\documentclass{article}

\usepackage{graphicx}
\usepackage[margin=1.4in]{geometry}
\usepackage{amsfonts}
\usepackage{amsmath}
\usepackage[english]{babel}
\usepackage{hyperref}
\usepackage{color}
\hypersetup{colorlinks=true,citecolor=blue,linkcolor=blue,filecolor=blue,urlcolor=blue}
\usepackage{tikz}
\usepackage{framed}
\usepackage{setspace}
\usepackage{nicefrac}
\usepackage[shortlabels]{enumitem}

\newcommand{\bv}{\begin{verse}}
\newcommand{\ev}{\end{verse}}
\newcommand{\be}{\begin{equation}}
\newcommand{\ee}{\end{equation}}
\newcommand{\bea}{\begin{eqnarray}}
\newcommand{\eea}{\end{eqnarray}}
\newcommand{\bq}{\begin{quotation}}
\newcommand{\eq}{\end{quotation}}

\newcommand*{\tr}{\mathrm{tr}\,}

\newcommand*{\ExpBegin}[1]{
\vspace{0.2ex}
\begin{center}
\begin{minipage}{\linewidth}
\begin{framed}
\vspace{-0.6ex}
{\centering {\bf #1} \\ }
\vspace{1.1ex}
\nobreak}

\newcommand*{\ExpEnd}{
\vspace{-0.8ex}
\end{framed}
\end{minipage}
\end{center}
}

\begin{document}

\title{QBism, Where Next?\bigskip}

\author{Christopher A. Fuchs\medskip
\\
\small Department of Physics, University of Massachusetts Boston
\\
\small 100 Morrissey Boulevard, Boston MA 02125, USA \smallskip
\\
\small and \smallskip
\\
\small JILA, University of Colorado, Boulder, CO 80309-0440, USA
}

\date{2 March 2023}

\maketitle

\begin{abstract}
This paper expresses what a breath of fresh air it has been since a few phenomenological philosophers have started to engage with QBism.  In service of the newfound discussion, the aim of this exposition is to lay out the structure of QBism as clearly as possible for that audience.  In the process, we arrive at eight tenets for QBism:  1) A quantum state is an agent's personal judgment.  2) A quantum measurement is an agent's action upon its external world.  3) Quantum measurement outcomes are personal to the agent performing the action.  4) The quantum formalism is normative rather than descriptive.  5) Unitary evolution too expresses an agent's degrees of belief.  6) Even probability-one assignments are judgments without ontic content.  7) Subjective certainty about what an outcome will be does not negate that unperformed measurements have no outcomes.  And, 8) quantum theory is a single-user theory for each of us.  We then analyze the Wigner's friend thought experiment in light of the eight tenets and indicate that a still more extended analysis is potentially QBism's surest path to uncovering an ontology to go with quantum theory's normative structure.  We conclude with a small discussion of how the philosophy of Maurice Merleau-Ponty may be relevant to this quest.
\end{abstract}

\tableofcontents

\section{Introduction}

\medskip

\begin{flushright}
\baselineskip=13pt
\parbox{3.1in}{\baselineskip=13pt
Hey, I'm just an old chunk of coal\\
But I'm gonna be a diamond some day\\
I'm gonna grow and glow 'til I'm so blue pure perfect\\
Gonna put a smile on everybody's face}\medskip\\
--- Billy Joe Shaver
\end{flushright}

\smallskip

QBism \cite{Fuchs10a,Fuchs13a,Fuchs14a,Fuchs2017} comes from humble beginnings.  It was not born fully formed as Botticelli's Venus was, nor as the Bohmian and Everettian interpretations of quantum mechanics purport to be.  Most strictly, QBism has always been a {\it research program}. Its long goal---to say something deep about the character of reality---was always at the top of the mind, but after 96 years of the quantum debate\footnote{Say, starting in 1927 with Heisenberg's uncertainty principle paper and Bohr's complementarity.}, a slow and careful methodology seemed called for. Less cheap, guesswork ontology\footnote{Is it a pilot wave? Is it parallel worlds? Does it consist of flashes of wave-function collapse? How about consistent histories? Maybe relations without relata? The full list of speculative ontologies that have been proposed is surely longer than can fit into any reasonably-sized footnote.}, more surgical dissection of the theory and an honest reckoning with what its structure has been trying to tell us all along. QBism's tack was to ask over and over, what is it about the world that makes us well-advised to use the calculus of quantum mechanics for structuring our probabilities?  Our {\it Bayesian\/} probabilities~\cite{Caves02a}.  Said this way, it became clear that the pertinent way to move forward was to get the ``epistemics'' of the theory right before anything else:  Getting reality right would follow for those who had patience enough to pass the marshmallow test~\cite{Mischel2014}.

In fact the first phase of QBism might be likened to a grand exercise in apophatic method: We won't yet tell you what reality is, but what it is not.  In particular, on the supposition that probabilities (even instances of probability-1) are not part of reality, after careful analysis, we'll tell you lots of other things that cannot be part of it either.  In this way, first the quantum state fell as a potential element of reality, then more surprisingly the operators used to describe quantum measurements, and then perhaps even shockingly Hamiltonians and unitary operators.  So it went with nearly every {\it individual\/} term of the theory.

In July 2002, after a frenzied year of applying the method more thoroughly than ever, I compiled a 229-page samizdat~\cite{Fuchs2002b} documenting how I was forced to my current position on these things by fighting tooth and nail with my colleagues Carlton Caves, David Mermin, and R\"udiger Schack. (They were all initially quite reluctant to go so far, and eventually Caves even jumped ship.) For the abstract of it, I used a passage written already in October 2001:
\bq
\noindent Collecting it up, it's hard to believe I've written this much in the little time since V\"axj\"o.  I guess it's been an active time for me.  I think there's no doubt that I've gone through a phase transition.  For all my Bayesian rhetoric in the last few years, I simply had not realized the immense implications of holding fast to the view that ``probabilities are subjective degrees of belief.''  Of course, one way to look at this revelation is that it is a {\it reductio ad absurdum\/} for the whole point of view, and this will certainly be the first thing the critics pick up on. But, you wouldn't have guessed less, I'm starting to view it as a godsend. For with this simple train of logic, one can immediately stamp out the potential reality/objectivity of any of a number of terms that might have clouded our vision.  With so much dead weight removed, the little part left behind may finally have the strength to support an ontology.
\eq
How prescient that passage was: For after 21 years, I don't remember even a single standard-style philosopher of physics agreeing with my perceived ``godsend.''  To their eyes the method of QBism left nothing whatsoever behind. I like the way my friend and QBism enthusiast Amanda Gefter~\cite{Gefter2014,Gefter2015,Gefter2022} once put it, ``Philosophers of physics will never accept the kind of reality QBism points them to because they can’t see it as a form of reality of any variety.''

So, what is that form of reality?  To this day, even QBists have the thinnest glimmer of it---but that is why QBism is a project, a research program.  The hints are strong that we will end up with something along the lines of what Will Durant once expressed so profoundly~\cite[p.\ 673]{Durant2006}:
\bq
\noindent The value of a [QBist pluriverse], as compared with a universe, lies in this, that where there are cross-currents and warring forces our own strength and will may count and help decide the issue; it is a world where nothing is irrevocably settled, and all action matters. A monistic world is for us a dead world; in such a universe we carry out, willy-nilly, the parts assigned to us by an omnipotent deity or a primeval nebula; and not all our tears can wipe out one word of the eternal script. In a finished universe individuality is a delusion; ``in reality,'' the monist assures us, we are all bits of one mosaic substance. But in an unfinished world we can write some lines of the parts we play, and our choices mould in some measure the future in which we have to live. In such a world we can be free; it is a world of chance, and not of fate; everything is ``not quite''; and what we are or do may alter everything.
\eq
But how to get there genuinely? That's where the real struggle still awaits us.  Pluriverse?  Unfinished world?  All action matters?  Everything is not quite?  What do all these things really mean?  How does one see them play out in the particular mathematical structure of quantum theory, and how can that structure {\it teach\/} philosophy something new to consider?\footnote{So that Bohr might revise (were he alive today) his opinion of philosophers: ``I felt that philosophers were very odd people who really were lost, because they have not the instinct that it is important to learn something and that we must be prepared really to learn something of very great importance.''~\cite{Bohr1962}}

QBism as it stands so far is only a scaffolding for building to those heights.  Key to everything though, is that it has done so with a scientific precision which general philosophizing cannot.  As I put it in~\cite{Fuchs2017}, ``One volunteers a philosophy, but one does not volunteer a physics.  A physics either flies in the world, or it falters and is eliminated by Darwinian selection.  That our most encompassing physical theory yet might lead to a philosophy once volunteered by temperament is a very powerful development.''

Yet, reciprocally, QBism was not born in a vacuum.  It has always needed inspiration from philosophy even in quite technical matters.  Perhaps the most direct example of this can be found in the quantum de Finetti theorem~\cite{Caves02b}, one of the first successes of the still-tentative QBist research program. This is because if one could prove a quantum de Finetti theorem, it would give a subjectivist Bayesian account of the phrase ``unknown quantum state'' in quantum state tomography~\cite{Paris2004}---a procedure that has often been flaunted as demonstrating that quantum states are objective after all.

Well, where did the idea of a quantum de Finetti theorem come from?  It came from our having recently learned in 1998 of the classic representation theorem of de Finetti himself from the 1930s.  In that case, the task at hand was to give a subjectivist account of the phrase ``unknown probability distribution.'' De Finetti understood that the existence of such a theorem would be essential to shore up his doctrine of ``probabilism''~\cite{deFinetti1989}. What few people realize, however, is that probabilism was not formulated in the service of statistical analysis, but was philosophy all the way down. As Richard Jeffrey wrote~\cite{Jeffrey1989}, ``For de Finetti the years before publication of {\it Probabilismo} (1931) were a time of explosive mathematical activity fired by his philosophical vision.''  In fact during its formation~\cite{deFinetti2006}, de Finetti was deeply under the influence of the Italian pragmatist movement of Giovanni Papini, Adriano Tilgher, Antonio Aliotta, and others of whom William James wrote so approvingly~\cite{James1978}. De Finetti himself confirmed in his autobiographical conclusion to {\it Probabilismo},
\bq
\noindent I found many things [in Burali-Forti and Mach] conforming to my ideas [but] there has recently been added a third and definitive base for my point of view: probabilism. It corrects and integrates the other two in the points that I could not accept:  {\it those in which anything seemed to be considered as having an absolute value, transcending the psychological value it has for me, and independent of it}. [my emphasis]
\eq
Thus the development of QBism was under the influence of pragmatism before anyone was even conscious of it!\footnote{Though, see my story in response to Question 14 of~\cite{Fuchs12} for an amusing follow-up.}

So without doubt, QBism has always needed help from philosophy and has on occasion gotten it.  It is just that little to none of the help ever came from standard philosophy-of-physics circles.  Philosophers of physics either wanted to make a fool of QBism\footnote{To be fair, some papers were at least honest efforts at it, as for instance~\cite{Myrvold2020}, but plenty more were just inane. Perhaps the most egregious of the latter was~\cite{Earman2019} by John Earman---a senior and quite respected man in the field---for it showed quite bluntly that he must not have read much of the QBist literature he cited~\cite{Fuchs2020}.} or at best temper its innovations until they could fit a semblance of them into a block-universe conception of nature seemingly so crucial to the milieu~\cite{Timpson08a,Bacciagaluppi2014, Glick2021,Fuchs2019}.  The years have taught us that, with a few exceptions,\footnote{See acknowledgements at the end of the paper.} there is probably not much reason to engage with that community further.

However, thankfully, standard philosophers of science do not exhaust the philosophical landscape relevant to physics.  As Berghofer and Wiltsche write in the introduction to this volume, ``The question is not whether QBists and phenomenologists should attempt to join forces, but what has taken us so long?''  To be sure, it is unlikely that QBism has exhausted the inspiration it might gain from a deeper plunge into the pragmatist writings of William James, John Dewey, F.~C.~S. Schiller\footnote{See for instance his monumental essay ``Axioms as Postulates''~\cite{Schiller1902}.}, and Richard Rorty, or other non-phenomenological philosophers who also took ``experience first'' such as Shadworth Hodgson~\cite{Hodgson1898} or Richard Avenarius~\cite{Lamberth1999}, but an unexpected workforce of eager phenomenologists is a most welcome development.  Who knows what might arise from this synergy?

The big thought on my own mind is to use our new collaboration to much deepen our understanding of the ``said form'' of reality that current QBism seems to indicate.  Will we land on Merleau-Ponty's ``flesh'' as the basic ontological element, instead of James's ``pure experience,'' Dewey and Bentley's ``transaction,'' or Whitehead's ``actual occasions?''  Or will it be still something else since none of those philosophies are responses to the particular mathematical details of quantum theory?  Whatever the outcome, I would love our final story to be {\it so blue pure perfect\/} that it may well lead us to the next stage of physics.  The volume collected here is our first step in that journey.

To that end, I will spend the better part of my contribution trying to give the most comprehensive exposition I can of where QBism stands {\it today}.  This way, all of us will be on the same page before proceeding to deeper contributions.  What exactly means QBism?  Indeed, I had already emphasized that QBism is an evolving project, but a newcomer may not be aware of how drastic the evolution has been---one should read Blake Stacey's contribution to this volume~\cite{Stacey2022} to see just how much so.  As he writes in his conclusion, ``Basically nothing posted on the arXiv before 2009 should be cited as an example of QBism, no matter who the authors are.'' That is, the mathematics is to be trusted, but maybe not the philosophy.  Nonetheless, even some of the most authoritative expositions since 2009 \cite{Fuchs10a,Fuchs13a,Fuchs14a,Fuchs2017,FuchsStacey2018,Mermin2019}---all important readings in their own right and necessary for giving texture to the project---often have different emphases than we commonly use now in our research groups at UMass Boston and Royal Holloway University of London.  Particularly, none of our writings to date have so emphasized the {\it normative aspects\/} of the mathematical structure of quantum theory.  This will hopefully be corrected here.

After Section~\ref{LoudScreechingHalt} (just described), which is the core of the paper, Section~\ref{Intermezzo} will reiterate how, though QBism may now have a firm hold on its ``eightfold path'' toward relieving the suffering at the ``quantum interpretations conventions,''\footnote{See the table on page 2 of Ref.~\cite{Fuchs2002}.} it is still in search of its ``noble truth'':  What is the precise ontology that compels the eightfold path?  In this regard, it currently strikes me that the best way forward will consist of deeper analyses of the Wigner's friend scenario~\cite{Wigner1961} and its extensions~\cite{Frauchiger2018,Baumann2020,Cavalcanti2021} than it has hitherto received.  Thus Section~\ref{Wigner's Hands} takes on Wigner's original argument from a QBist perspective to lay some groundwork for that future discussion.  What becomes clear is that in QBism, Wigner and his friend must be treated symmetrically, much like Merleau-Ponty did with his hand argument.  The paper concludes in Section~\ref{DuhDuhDuh} with some hopeful remarks on what we might squeeze out of such further analyses.

\section{QBism, Where Currently?}  \label{LoudScreechingHalt}

\medskip

\begin{flushright}
\baselineskip=13pt
\parbox{3.1in}{\baselineskip=13pt
I'm gonna learn the best way to walk\\
I'm gonna search and find a better way to talk\\
I'm gonna spit and polish my old rough-edged self\\
'Til I get rid of every single flaw}\medskip\\
--- Billy Joe Shaver continues
\end{flushright}

\medskip

In Subsection~\ref{KikiMowing} I first describe what is meant by the terms ``agent'' and ``user of quantum mechanics'' from a QBist perspective.  (Boy, the phenomenologists might really help us here! The shortcomings will be quite apparent.) I then devote the remaining Subsections~\ref{DutchieBoy} to~\ref{SingleUser} to the much more developed parts of QBism, explicating eight of its key tenets in significant detail.

\subsection{Agents and Users of Quantum Mechanics}
\label{KikiMowing}

For QBism, the quantum formalism is a tool decision-making agents are advised to adopt in light of the peculiar uncertainties we find in our world. Namely, the theory guides its users in how to better gamble on the personal consequences (``experiences'' or ``lived experiences'') of their actions on physical systems. Particularly in QBism, the quantum formalism plays a {\it normative\/} role for its users; it does not play a descriptive role concerning exactly how the world is. Its focus is on how a user {\it should\/} gamble.

But then what is a ``user of the theory?'' In the following, we will make a distinction between agents broadly speaking and the users of quantum mechanics:
\begin{itemize}
\item
An {\it agent\/} is an entity that can freely take actions on parts of the world external
to itself and for which the consequences of its actions matter to it.
\item
A {\it user of quantum mechanics\/} is an agent who applies the
quantum formalism normatively for better decision making.
\end{itemize}

While our definition of a user is narrow, our definition of an agent is broad.  An agent is any part of the world that can act autonomously on other parts of the world and can be analyzed fruitfully in teleological terms.  Hence the phrase {\it matter to it}.  Our definition does
not rule out dogs, euglena, or even artificial life (if there can be such a thing) as agents.  However, it does exclude a computer hard wired from the outside to
deterministically ``choose'' its actions from a look-up table (essentially a Turing machine).  It also excludes electrons: For though an electron may act autonomously on its external world, it is hard to think that the consequences of its actions matter to it.

On the other hand with regard to the notion of a user of quantum mechanics, as Khrennikov emphasized of QBism in Ref.~\cite{Khrennikov17}, ``The idea is that QM is something used only by a privileged class of people. Those educated in the methods of QM are able to make better decisions (because of certain basic features of nature) than those not educated in the methods of QM.'' By this light, Werner Heisenberg was an agent in 1924, and likely even a user of probability theory, but he was not yet a user of quantum mechanics.  He himself did not enter that privileged class until 1925-1926.  Will IBM ever be able to construct an entity without DNA, one made of silicon and copper, that would count as a user of quantum mechanics?  Maybe.  Maybe not.  But that is irrelevant to the view of quantum theory as an addition to decision theory.  To put it in a slogan, ``Quantum mechanics is a user's manual ready and waiting for anything that can make use of it.''

Hereafter, we will assume most of the agents we are speaking of are in fact users of quantum mechanics, but not always.  So please stay aware of the context.

There certainly exists a range of definitions of agency in the literature, some overlapping with our definition to different degrees, some not remotely recognizable from our point of view (see any of the works of Daniel Dennett for instance). For comparison or contrast on how another physicist influenced by the phenomenological tradition---this time in the form of Martin Heidegger---has tackled the issue, see Refs.~\cite{Mueller2018,Briegel2012}.

One very interesting issue still in need of greater fleshing out is this.  Suppose a team of scientists sharing notebooks, calculations, observations, etc., can be considered an ``entity freely taking actions on its external world.''  Then according to the above definitions it can count as a single agent and even a user of quantum mechanics~\cite{DeBrotaStacey2018,DeBrota2020b}.  Why shouldn't Napoleon's Grand Arm\'ee count as a single agent for some purposes?  Why shouldn't the experimentalists and theorists of the first-ever continuous-variable quantum teleportation experiment~\cite{Furusawa1998} together count as a single user of quantum mechanics?

Or even backing off from quantum mechanics per se, consider this elegant example recently put forth by Jacques Pienaar~\cite{Pienaar2022}:
\bq
\noindent Think of a highly trained volleyball team. The ball appears: who will take it? Where will they aim to hit it? For a team that is sufficiently cohesive, these decisions will be made without explicit speech. Subtle cues like bodily stances and small movements may contribute to indicating who should act, and how. Such movements may only be perceived subconsciously, and decisions made intuitively. The feeling would be that each member of the group knew that ``that was the right move'', yet nobody could say in which one of their minds the idea originated, and nor did it have to be explicitly communicated from one mind to all the rest. The same thought arose in all of them, all experienced it, but it is experience as a thought that belongs to none of them in particular, and the thought could only arise when they are all together in that situation.
\eq

Nonetheless, there may be unforeseen nuances in this way of thinking that {\it may or may not\/} contradict other bits and pieces of QBism.  Surely more needs to be said!  Perhaps the issue is in part captured by what Cavalcanti~\cite{Cavalcanti2021} calls a ``Wigner bubble,'' but we leave it as a tantalizing possibility for one of the many technical directions still to be explored.

\subsection{A Quantum State Is an Agent's Personal Judgment}
\label{DutchieBoy}

In QBism, the {\it exclusive\/} purpose of the quantum formalism is to help an agent make better decisions.\footnote{As an aside, one might argue that this is the best sense of ``explanation'' as well~\cite{Fuchs2015}.}  That may sound instrumentalist, but it is not.  It only means that if something is to be inferred about reality from the formalism, then it has to be done in the way of the archeologist:  Why did this civilization design this tool for the terrain it lived in? What are the extant conditions that made {\it quantum theory as a tool\/} the natural choice for agents to gamble best with reality?
With regard to this way of putting it, QBism excels above any of its sci-fi competitors~\cite{FuchsInterview}, which really don't care where the formalism came from.
History has shown that the rigorous use of the quantum formalism enables an agent to make more successful gambles in navigating the world than he would have otherwise.  Why?  It's QBism's take that when an answer is found we will finally understand John Wheeler's ``How come the quantum?''~\cite{Wheeler1987}. But the first step toward the goal is to get straight what quantum theory is actually about.  This is the reason for QBism's unflinching stance that the exclusive purpose of the quantum formalism is to help an agent make better decisions.

It is unfortunate that the term gamble evokes images of games of luck, but we use it in a sense that is meant to encompass any action an agent can take where the consequences matter to the agent. Any physics experiment is thus a gamble.  As we will explain in more detail in Section~\ref{NormativitySection}, the quantum formalism can be viewed as an addition to classical decision theory~\cite{Fuchs13a,Fuchs2017}.
Particularly, following the approach to decision theory pioneered by Bruno de Finetti, Frank P. Ramsey, and L. J. Savage \cite{Savage1972,BernardoSmith1994}, QBism takes all probabilities to be specific to the agent using them---they are personal, quantified degrees of belief.  And when the subject matter is quantum mechanics, the probabilities involved are an agent's personal degrees of belief concerning their future measurement outcomes.

Personalist probabilities \cite{deFinetti1990,Berkovitz2019} acquire an operational meaning by their use in decision making. A simple case of these considerations gives the so-called Dutch-book argument for the probability calculus.  There, an agent's ``probability'' $P(D)$ for an event $D$ is identified with her {\it valuation\/} of a lottery ticket which pays \$1.00 if $D$ occurs and \$0.00 otherwise. A ticket like this is said to have a valuation of $x$ if the agent (privately, perhaps secretly) commits herself to the following: Whenever offered a ticket, she will buy it for any amount less than $\$x$, and whenever asked to sell, she will do so for any amount offered larger than $\$x$.  Similarly, one can speak of a ``conditional probability'' $P(D|H)$ in terms of the valuation of a lottery ticket for a compound event.  In this case,  $\$P(D|H)$ corresponds to the threshold price for a ticket that returns \$1.00 if both $H$ and $D$ occur, but if $H$ does not occur, all transactions are returned to both buyer and seller.

What are the ``correct'' valuations for these tickets?  Clearly they can only depend upon what the agent believes about the events $H$ and $D$---it is about the agent's money after all and the risk she is willing to take.  Most importantly, the valuations are not something determined by the agent's external world, but are genuinely personal, in some measure corresponding to her own autonomy.\footnote{There is a wonderful dialog in the movie {\sl The Ballad of Buster Scruggs\/} directed by Joel and Ethan Coen that captures this idea perfectly. The setting is that of five people squeezed into a stagecoach heading to a certain Fort Morgan. The key part of dialog is between a Frenchman (Ren\'e) and a morally upright Lady (Mrs.\ Betjeman):

LADY:  I have been living with my daughter and son-in-law these last three years.

FRENCHMAN:  Ahh, the parent should not burden the household of the child.  This was wrong of you, {\it madame}.

LADY:  I was not a burden!  I was \underline{welcome} in my daughter’s house!

FRENCHMAN:  Ahh, she would say so, of course.  But no doubt you could read in her facial expressions \ldots\ that your presence was not wanted.  We each have a life.  Each a life, only our own.

LADY:  You know nothing of me or my domestic affairs!

FRENCHMAN:  I know that we must each spin our wheel, play our own hand.  I was once at cards with a man named Cipolski.  This was very many years ago \ldots\ We were at cards.  My hand was poor; I folded.  But Cipolski and four others remain.  Cipolski said to me, Ren\'e I am in distress, you must play for me while I perform {\it mes n\'ecessit\'es}---ah, my necessaries.  ({\it Lady giggles uncomfortably.})  I say, friend, no I cannot wager for you.  He says, but of course you can, we know each other well---you wager as I would do.  I say, it is quite {\it impossible\/}, no?  How a man wagers, it is decided by who he is, by the entirety of his relation to poker right up until the moment of that bet.  I cannot bet for you.  {\it Pourquoi pas}?  I cannot \underline{know} you.  Not to this degree.  We must each play our own hand.  No, Cipolski, I say.  No, we may call each other friend, but puh we cannot know each other so \ldots

LADY:  Poker is a gambling game.  ({\it Frenchman nods.})  You have pursued a life of vice and dissipation---and you are no doubt expert in such pursuits.  But no conclusions drawn from such an existence will apply to a life rightly lived.

FRENCHMAN:  Life is life.  Cards will teach you what you need to know. \ldots

}  The last thing one would want to say is that the valuations are somehow properties of the events $H$ and $D$ themselves or even the lottery tickets.

Further, note that the word ``probability'' as associated with these lotteries is {\it so far\/} a mere placeholder:  It is a notion mildly evocative of how probability is used, but we might have called it by any other name since there is so little structure.  As it stands, there is no mathematical specification to the merely symbolic $P(D)$.

Remarkably, however, the full structure of probability theory can in fact be derived by adding one simple normative requirement to an agent's valuations:  That whatever assignments she makes for $P(H)$, $P(D)$, $P(D|H)$, etc., they should never be such that there exists a strategy of buying and selling which leads to a {\it sure loss\/} for the agent---i.e., a net loss for the agent no matter which outcomes occur.  If such a strategy exists, then one says that a ``Dutch book'' can be made against the agent.  If a Dutch book cannot be made, then one says that the agent is {\it Dutch-book coherent}, or simply {\it coherent}.  Thus, when Dutch-book coherence is satisfied, one has every right to call these lottery valuations probabilities in the proper sense of mathematical probability theory.

More specifically, from the requirement of coherence, it follows that valuations must be nonnegative and bounded, $0\le P(H) \le 1$ and $0\le P(D|H) \le 1$, etc.  When $H$ and $D$ are mutually exclusive, valuations must be additive
\be
P(H,D) = P(H) + P(D)\;.
\ee
Bayes' rule must be satisfied,
\be
P(H)P(D|H)=P(D)P(H|D)\;,
\ee
and similarly for all the more elaborate statements of probability theory.\footnote{Or at least probability theory over finite sets of events.}

One such more-elaborate statement which arises directly from Dutch-book coherence is the Law of Total Probability (or LTP)\@.  Since the LTP will play a significant role in our later discussions, it is worth expressing it in the present context.  Consider an agent who contemplates two sets of events ${\cal R}=\{R_1, R_2, \ldots, R_n\}$ and ${\cal E}=\{E_1, E_2, \ldots, E_m\}$, each of which is mutually exclusive within itself. (We use distinct indices $n$ and $m$ because there is no need for the sets to have the same cardinality.)  Taking into account all the lotteries and compound lotteries which can be formed for these events, the agent derives that in order to be coherent she must satisfy
\be
\label{Gizmo}
P(E_j) = \sum_{i=1}^n P(R_i) P(E_j|R_i) \qquad \forall j\in\{1,2,\ldots,m\}\;.
\ee
Note one thing about this statement:  In its very set-up, it is assumed that both the ${\cal R}$ event and ${\cal E}$ event will come to be recognized---so that all lottery tickets can either be returned or paid off appropriately.  This will be an important point when our discussion turns to the role of the Born rule in quantum theory in Section~\ref{NormativitySection}.

So much for characteristics of personalist probability theory in the most general setting.  The way this makes a connection to quantum theory is through the fact that any quantum state can be identified with an ``expectation catalog'' for the outcomes of all possible quantum measurements---a point first emphasized by Schr\"odinger in 1935~\cite{Schroedinger1935}.  QBism's strategy has thus been to understand probability first and quantum mechanics next, as directed by E.~T. Jaynes~\cite{Jaynes1990}.  After trials with other potential meanings for probability in the 1990s (as frequencies, propensities, objective chances, etc.), the originators of QBism ultimately settled on the idea that the only consistent and operationally meaningful interpretation of probability is the personalist Bayesian one. Consequently a quantum state, from the point of view of QBism, must be understood as a catalog of personal expectations if it is to be anything meaningful at all.

To put it succinctly, where Bruno de Finetti declared ``probability does not exist'' to convey the idea that probabilities cannot be considered properties of the agent's external world \cite{deFinetti1990,Berkovitz2019}, QBism declares ``quantum states do not exist'' in the same sense.  Just as all probabilities are personal judgments, within QBism all quantum states are personal judgments. Of course, each probability assignment in the catalog must be Dutch-book coherent, but what is interesting in quantum theory is the way in which all these Dutch-book coherent judgments are implicitly tied to each other.  We will explain this in more detail in Sections~\ref{CaptainQBism} and \ref{NormativitySection}.  The lesson for the moment though is that a quantum state is a personal judgment and not something mandated by the world external to the agent.

\subsection{A Quantum Measurement Is an Agent's Action Upon Its External World}
\label{CaptainQBism}

A measurement is an action of an agent on its external world, where the consequences of the action (its {\it outcomes} in more usual
terminology) matter to the agent.  They should matter enough that the agent would be willing to gamble upon them.

Like our definition of agent before, this definition of measurement is very broad. Basically anything an agent can do to its external world---from opening a box of cookies, to crossing a street, to performing a sophisticated quantum optics experiment---counts as a measurement in our sense. The only thing that sets a quantum measurement (as normally construed) apart from a more pedestrian example is whether it is fruitful or worth one's while to apply the quantum formalism in analyzing it.  In many situations, there is little to be gained by analyzing the consequences of one's actions through the aid of raw probability theory, much less the full-blown apparatus of the quantum formalism.  Think of crossing a street: Already one has an intuitive-enough feel for when and when not to cross that pausing to make a calculation would be self-defeating.  However, there are some situations where it is absolutely crucial to invoke the quantum formalism if one wants to try to be maximally prepared for nature's consequences. One such case is obviously the quantum optics lab, but another, more extreme case concerns an agent's actions on another living system when treated with the full generality of quantum theory. This, for instance, is the case for the Wigner's friend thought experiments, where the usage of quantum theory is pushed to its most extreme, and perhaps most revealing, analysis.

Let us work toward developing what we mean by applying the quantum formalism to analyze one's gambles.  First, consider two general actions ${\cal E}$ and ${\cal E}^\prime$ an agent might take on some object in his external world.  For instance, the agent might be a boxer in training who is very carefully watching and analyzing his practice games before going out for a real match-up.  The two actions contemplated might be aiming a right cross or a left hook at his opponent.  Throughout, we will implicitly identify the actions an agent can take with the sets of potential experiences or consequences they will lead to for him.\footnote{Note the similarity between this notion and the way R.~T. Cox, the founder of E.~T. Jaynes's preferred school of Bayesianism, models the notion of a {\it question}: A question is the set of all its possible {\it answers}~\cite{Cox1961}.}  Thus, ${\cal E}$ and ${\cal E}^\prime$ can be expanded as ${\cal E}=\{E_1, E_2, \ldots, E_n\}$ and ${\cal E}^\prime=\{E_1^\prime, E_2^\prime, \ldots, E_m^\prime\}$, where the $E_i$ and $E_j^\prime$ represent, for instance, the {\it mutually exclusive\/} potential subsequent jabs the boxer could receive from his opponent in return for his initial choice of ${\cal E}$ or ${\cal E}^\prime$.  (Again, there is no need that the cardinality of the two sets be the same.)  To say that an agent will gamble on the consequences of these actions means that he must settle his mind on probability distributions for the consequences of each:  Sets of nonnegative numbers $P(E_i)\ge 0$ and $P(E_j^\prime)\ge 0$ such that
\be
\label{SoongChi}
\sum_{i=1}^n P(E_i)= 1 \qquad \mbox{and} \qquad \sum_{j=1}^m P(E_j^\prime)= 1\;.
\ee
Through these numbers, the agent may then use decision theory to help make his choices of when to take which action, and this is where a normative theory for guidance plays its role.  If the boxer incorporates the lessons learned from practice into his split-second thinking he will be much better prepared for his game.

The same is true of any two actions that might be considered in any setting.  However, note that in the case of the boxer it is essentially impossible for him to perform both a left hook and a right cross at the same time.  This is reminiscent of the complementarity one sees in some quantum measurements, like position and momentum.  With this in hand, we are ready to be more precise about the meaning of ``applying the quantum formalism.''

To apply the quantum formalism means to ask for further guidance than unadorned probability theory, as represented by Eq.\ \eqref{SoongChi}, can supply.  This is because probability and decision theory take into account no details of the worlds in which they are used:  Their edicts are independent of the characteristics of the worlds their agents inhabit.  Quantum theory on the QBist view, however, is an {\it addition\/} to probability theory which very much takes into account the unique characteristics of our given world.  If our world were a different world, agents would not be well advised to use the quantum formalism.

This is to say, to apply the quantum formalism an agent decides it is to his benefit to associate a Hilbert space with the object he intends to act upon and, by one means or another, establish an association
\be
\label{MathematizeMyFeelings}
E_i\; \longleftrightarrow\; \hat E_i \qquad \mbox{and} \qquad E_j^\prime\; \longleftrightarrow\; \hat E_j^\prime \qquad \forall\; i,j
\ee
between his expected experiences and the elements of some positive-operator valued measure (POVM) on the Hilbert space.  (Note the hats on the right-hand symbols to denote that they are operators, rather than direct expressions of the experiences.) This means sets of operators $\big\{\hat E_i\big\}$ and $\big\{\hat E_j^\prime\big\}$ such that
\be
\langle\psi|\hat E_i|\psi\rangle\ge0 \qquad \mbox{and} \qquad \langle\psi|\hat E_j^\prime|\psi\rangle\ge0 \qquad \forall\; i, j, \mbox{ and } |\psi\rangle\;,
\ee
and
\be
\sum_i \hat E_i = \hat I \qquad \mbox{and} \qquad \sum_j \hat E_j^\prime = \hat I,
\ee
where $\hat I$ is the identity operator on the Hilbert space.\footnote{In this notion of quantum measurement, there is no requirement that the operators $\hat E_i$ be orthogonal to each other. Nor are the cardinalities of the sets ${\cal E}$ and ${\cal E}^\prime$ limited by the dimension of the Hilbert space.  This is because the $\hat E_i$ are formal tools for associating probabilities with the agent's mutually exclusive experiences $E_i$. This may be a point of confusion for those in the philosophy of physics community who are more familiar with the von Neumann notion of measurement or with propositional concepts drawn from the tradition of quantum logic.}

POVMs are well-known and essential to quantum information science, but appear to be not so familiar to philosophers of physics. The key reason QBism adopts this notion of ``applying the quantum formalism'' is that POVMs represent the most general kind of measurement one can perform in quantum mechanics~\cite{Nielsen2010}.  They therefore can model any action an agent can take upon its external world.  Moreover, without POVMs (i.e., restricting to von Neumann measurements alone\footnote{As Earman's imagined QBians do~\cite{Earman2019,Fuchs2020}.}), QBism's mathematical project of rewriting quantum theory to make its normative content manifest would be stymied. There will be much more to say on this in Section~\ref{NormativitySection}.

Once this much is done, the way quantum theory completes its normative guidance is by suggesting the agent strive to find a {\it single\/} quantum-state assignment $\hat\rho$ (pure or mixed) such that his declared probabilities $P(E_i)$ and $P(E_j^\prime)$ may be calculated according to the Born rule:
\be
\label{FirstBornSon}
P(E_i)=\tr\hat\rho \hat E_i  \qquad \mbox{and} \qquad P(E_j^\prime)=\tr\hat\rho \hat E_j^\prime\;.
\ee
What this expresses is that the agent's {\it probability assignments\/} for the outcomes of various hypothetical measurements (or even, say, the performed and unperformed experiments Asher Peres contrasted in~\cite{Peres78}) should not be loose and unhinged from each other.  Similarly so with three hypothetical measurements, four hypothetical measurements, or any number.  Indeed, when an agent contemplates how he will gamble upon the outcomes of one imagined measurement, he ought to take into account how he will gamble upon the outcomes of all others he might imagine.

Notice how this differs from a more common presentation of probabilities within quantum mechanics.  There, the quantum state $\hat\rho$ is almost always treated as something sacrosanct, and probabilities are {\it derived from it}.  Where the specification of the quantum state comes from in the first place is usually unquestioned---for instance, it is often simply supposed as in so many textbook exercises---but here the tables are turned. A quantum state, rather than having a status logically prior to the probabilities in Eq.\ \eqref{FirstBornSon}, is something practically {\it born\/} with them, as the agent tries his best to take into account the peculiarities of the quantum world.

The normative thrust of quantum theory is that it is a kind of glue for probability assignments over and above the requirements of raw probability theory.  From the QBist perspective, it is this glue which indicates the physical content of quantum theory.  The world is such that agents should adopt this extra requirement on their gambles.

But what if given an agent's assignments $P(E_i)$ and $P(E_j^\prime)$, there is no $\hat\rho$ such that Eq.\ (\ref{FirstBornSon}) is satisfied?  Then it means the agent should rethink why he believes what he believes.  Perhaps his mapping from experiences to POVM elements in Eq.\ (\ref{MathematizeMyFeelings}) should be rethought.  Perhaps he should rethink the dimensionality of his chosen Hilbert space for the system.  Perhaps he should simply rethink whether he really wants to assign $P(E_i)$ and $P(E_j^\prime)$ in the first place.  Do these valuations genuinely fit with his larger mesh of beliefs beyond those for the outcomes of the two contemplated measurements?  Maybe further thinking or detailed calculation should be made before the bold leap of writing down $P(E_i)$ and $P(E_j^\prime)$. However, on exactly what needs to be adjusted, quantum theory gives no guidance. It merely indicates that {\it something\/} needs to be adjusted, and this is what it means that the quantum formalism is a normative suggestion rather than a direct description of reality.  Again, we will expand on this in Section~\ref{NormativitySection}.

One final thing for this section:  Note that we were careful to apply the term measurement only to actions an agent takes on his external
world.  We thus require a strict separation between the agent performing the measurement and the system this measurement is an action upon.
Why would this be?  Why make such a restriction?  In fact, it is no restriction at all, but a logical requirement for our setting.  What could
it mean for the agent to take actions upon himself without conceptually backing off from the very distinction between the agent (the
autonomous seat of action) and its external world we have taken as our starting point? The two would have to be identified after all,
contrary to the very spirit of QBism which aims to replace the notion of a block universe with a Jamesian
pluriverse~\cite{Fuchs2017,Fuchs12,Samizdat2,Fuchs16,Fuchs18}.

Furthermore this separation means that there is no sense in QBism to an agent assigning a
quantum state $\hat\rho$ to himself.  This follows from the fact that there are no normative conditions of the form Eq.\ \eqref{FirstBornSon}
to try to satisfy.  The quantum foundations literature is rife with discussions of agents assigning quantum states to themselves, but in QBism it is a {\it contradictio in terminis}.

\subsection{Measurement Outcomes Are Personal to the Agent}

QBism owes much of its development to the influence of Wolfgang Pauli \cite{Fuchs2010}, but this is one idea of Pauli's that QBism ultimately had to disabuse itself of:
\begin{quote}
[T]he objectivity of physics is fully preserved in quantum mechanics in the following sense.  Although according to the theory it is in principle only the statistics of test series that are determined by law, the observer cannot influence the result of his observation---such as the response of a counter at a certain moment---even in the unpredictable individual case.  The personal characteristics of the observer are in no way included in the theory; rather, the observation can be carried out using objective recording devices, the results of which are objectively available for everyone's inspection.
\cite[p.\ 122, but improved with Google Translate]{Pauli94}
\end{quote}
The first part of this relies on a frequentist understanding of probability, which is disappointing, but it is the last sentence that is particularly troublesome for QBism, as it was a block to our clear thinking for quite a number of years.  With hindsight, that QBism would ultimately reject the notion of ``results of which are objectively available for anyone's inspection,'' was an inevitable consequence of QBism's long progression from the personalist Bayesian notion of probability to the subjectivity of quantum states to the personal nature of measurement operators and the realization that unitary evolution itself is a personal judgment~\cite{Fuchs2002}.  (Also see~\cite[Introduction]{Samizdat2}.) The final straw came from contemplating how measurement operators could be subjective judgments at the same time as having results that obtain the same meaning for anyone who sees the outcomes.  The tension was insurmountable, and QBism ultimately opted to take more seriously a different utterance of Pauli's: ``[I]t is allowed to consider the instruments of observation as a kind of prolongation of the sense organs of the observer \ldots''~\cite{Pauli55}.  QBism takes this phrase to its logical conclusion. Whereas Pauli, Bohr, and the other Copenhageners always had their classically describable measuring devices mediating between the registration of a measurement's outcome and the observer, for QBism the measuring instrument was taken to be literally part of the agent and the measurement outcome his or her direct experience.\footnote{See the recent paper of J. Pienaar~\cite{Pienaar2020} for a much fuller explication of this than anything written before.}

When an agent performs a measurement---that is, takes an action on its external world---the ``outcome'' of the measurement is the consequence of
this action for his or herself. A measurement outcome is personal to the agent doing the measurement.  Thus two agents cannot strictly speaking experience the same outcome. Different agents may inform each other of their outcomes and thus agree upon the consequences of a measurement, but a measurement outcome should not be viewed as an agent-independent fact which is available for anyone to see~\cite{Caves07}.

This tenet has led some commentators to claim mistakenly that QBism is a form of solipsism. This claim has been thoroughly refuted, not only by  the QBists themselves~\cite{Fuchs10a,DeBrotaStacey2018,Fuchs16,Mermin2014} but even by some professional philosophers~\cite{Timpson08a,Glick2021,Healey2016,Wallace2016}. Among other reasons, QBism is not solipsism simply and immediately because of the premise that a measurement is an action on the {\it world external to the agent}. A QBist assumes the existence of an external world from the outset.  Furthermore, the consequences of measurement actions are beyond an agent's control---the world can surprise the agent---as Pauli himself noted in his discussion of objectivity. The external world is thus capable of exhibiting genuine novelty in response to an agent's actions---i.e., the world and the agent cannot be identified with one another. (See~\cite[pp.\ 6--10]{Fuchs16} and \cite[pp.\ 19--20]{Fuchs2017}, {\tt arXiv.org} versions.)  The entire basis for calling QBism solipsism is just short-circuited by the concepts QBism relies upon for its very starting point.

\subsection{The Quantum Formalism Is Normative Rather Than Descriptive}
\label{NormativitySection}

We have already said much about how in QBism the quantum formalism is understood normatively, rather than descriptively as it is in nearly every other quantum interpretation.\footnote{Perhaps the only exception is Richard Healey's ``pragmatist interpretation'' of quantum theory~\cite{Healey2017}, but it is so very far from QBism in what it takes to be a norm---for instance, it even posits a ``correct'' quantum state for a given perspective as a norm~\cite{Healey2022}---that it essentially shares nothing in common with QBism except the word ``normative.'' Moreover, it is simply a wholly different interpretative framework: Quantum mechanics could be posed even if there were never an agent in sight, and sin of all sins, it supports a block-universe picture of nature!}  Yet, this can be argued still more convincingly through a detailed analysis of the Born rule.  In our discussion surrounding Eq.\ (\ref{FirstBornSon}), the topic was the gluing together of probabilities for the outcomes of distinct quantum measurements. However, even for a single measurement the Born rule should be viewed as placing additional constraints on an agent's probability assignments---extra constraints upon which probability theory is simply silent.  Indeed QBism sees the Born rule as the most fundamental addition to probability theory that the quantum world hands its agents.  To understand the quantum formalism, one must first understand the Born rule.

As in Section~\ref{CaptainQBism}, we consider the expression of the Born rule as it applies to the most general notion of quantum measurement:  A positive operator-valued measure, or POVM, $\big\{\hat{E}_1, \hat{E}_2, \ldots, \hat{E}_n\big\}$.  The agent's probability $Q(E_j)$ for the experience $E_j$, formally expressed by a positive semi-definite operator $\hat E_j$, is given by
\be
\label{SecondBorn}
Q(E_j) = \tr \hat\rho \hat E_j\;,
\ee
where $\hat\rho$ is some density operator.  This time we use the notation $Q(E_j)$ to call attention to the fact that our probability
assignment comes from a quantum mechanical calculation. However just as before, the operator $\hat\rho$, the distribution $Q(E_j)$, and the
association
\be
\{E_1, E_2, \ldots, E_n\} \quad\longleftrightarrow\quad \big\{\hat{E}_1, \hat{E}_2, \ldots, \hat{E}_n\big\}
\ee
are all to be understood as
subjective judgments.  If for whatever reason the chosen $\hat\rho$, the
$Q(E_j)$, and the $\hat E_j$ do not satisfy Eq.\ (\ref{SecondBorn}), then the guidance of quantum theory is that the agent should modify at
least one of the terms.  This guidance however does not prescribe which of the terms to modify or how to modify them; none of the terms should be considered as having logical priority over any of the others.  Indeed, all the terms live at the same conceptual level.  In this way quantum theory's role is analogous to the kind of guidance Dutch-book coherence gives:  It speaks of the recommended relations between the terms, but nothing of their exact values, the latter being the provenance of the agent using the theory.

This is QBism's stance, but it is one thing to declare it, and another to make it compelling to one's readers.  Why, unless one has already drunk the Kool-Aid, should $\hat\rho$, the $Q(E_j)$, and the $\hat E_j$ all live at the same conceptual level?  Moreover, our whole description begs a question: If the POVM $\big\{\hat{E}_1, \hat{E}_2, \ldots, \hat{E}_n\big\}$ is a personal judgment, what exactly is it a judgment of?  These questions can be most easily answered by introducing the notion of a ``reference apparatus'' for a given quantum system.\footnote{Introducing a {\it reference measurement\/} is a conceptual technique with a long history in QBism, going back to the proof of the quantum de Finetti theorem~\cite{Caves02b} so crucial for analyzing the notion of an ``unknown quantum state'' from a QBist perspective.  The notion of a reference apparatus is the same concept but with the addition of a set of post-measurement quantum states for the system.}

A {\it reference apparatus\/} for a $d$-dimensional quantum system is {\it any\/} positive operator valued measure $\big\{\hat{R}_i\big\}$ with $d^2$ linearly-independent measurement operators, along with an associated set of linearly independent post-measurement quantum states $\{\hat\sigma_i\}$.  One can prove that there are infinitely many such structures for each $d$.  However, the word ``reference'' refers to the requirement that once a choice of one of these is made, it must stay fixed for all the agent's calculations---this is like choosing a reference frame in relativity theory.  What is significant about such an apparatus is that any operator can then be written as a {\it unique\/} linear combination of either of the two sets. This follows because each set of operators form a complete (though nonorthogonal) basis for the $d^2$-dimensional vector space in which $\hat\rho$ lives.  Moreover the expansion coefficients for $\hat\rho$ in terms of the $\hat{R}_i$ are determined by none other than the Born rule probabilities $P(R_i) = \tr \hat\rho\hat{R}_i$ themselves.  This follows because $(\hat A,\hat B)=\tr \hat A^\dagger B$ satisfies the properties of an inner product for the space.

This sets the stage for thinking of a quantum state as not only an expectation catalog for the outcomes of all possible measurements (Schr\"odinger again), but as a {\it single\/} probability distribution $P(R_i)$ full stop.  With such an identification
\be
\hat\rho \quad\longleftrightarrow \quad  P(R_i)\;,
\ee
one sees that it is a matter of course for a QBist to treat a quantum state as subjective as any personalist Bayesian probability distribution: For then a quantum state {\it just is\/} a probability distribution---it has no further content above and beyond that (contra~\cite{Barzegar2020} to some of QBism's critics~\cite{Brown2017}).

Indeed, this mapping gives a means to rewrite the entire Born rule in purely probabilistic terms.  To see this, consider the following scenario.  An agent has a physical system for which she plans to carry out either one of two mutually exclusive actions or protocols on it.  In the first protocol, she imagines measuring the system directly according to the operators $\big\{\hat{E}_j\big\}$ of any general POVM and thereby obtaining some outcome $j$.  However in a second, alternative protocol, she imagines cascading two measurements---a kind of one-two punch---first performing the reference apparatus and only subsequently performing the POVM $\big\{\hat{E}_j\big\}$ of the first protocol.  In this case, she would obtain two outcomes $i$ and $j$, not one.

\begin{figure}
\begin{center}
    \includegraphics[width=4.5in]{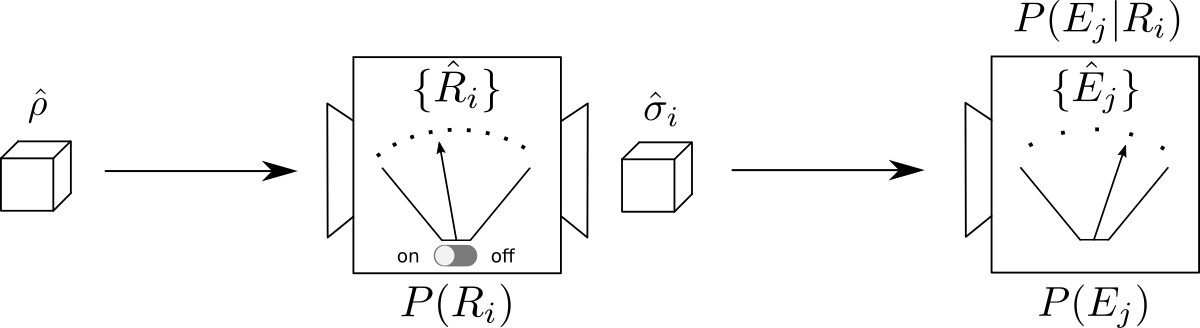}
\end{center}
\begin{center}
    \includegraphics[width=4.5in]{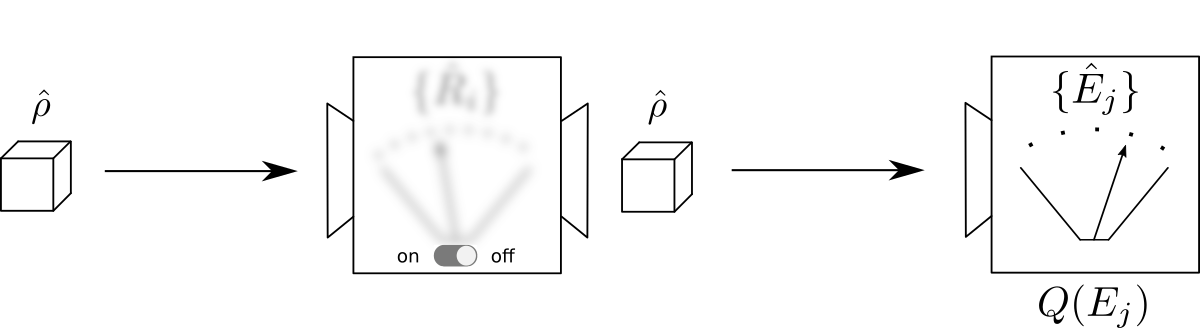}
\end{center}
\caption{Two distinct experiments.  In QBism, the Born Rule is not about either one individually, but rather about the connections between their probabilities.  In the top experiment, the reference device is turned on so that there are three probabilities in its telling: $P(R_i)$, $P(E_j|R_i)$, and $P(E_j)$. They must satisfy the Law of Total Probability, Eq.~(\ref{Gizminko}). However, in the bottom experiment the reference device is turned off---there is only one probability $Q(E_j)$ in its story. The Born Rule is the narrative glue that ties the two stories together.}
\label{Metzenbaum}
\end{figure}

It must be noted, probability theory in the abstract \emph{provides no consistency conditions} between the agent's expectations for the outcomes of these two protocols.  Let $P$ denote all of the agent's probability assignments for the consequences of following the two-step protocol and $Q$ those for the single-step protocol.  Then there is no doubt that the $P(E_j)$, $P(R_i)$, and $P(E_j|R_i)$ must satisfy the Law of Total Probability,
\be
\label{Gizminko}
P(E_j) = \sum_{i=1}^{d^2} P(R_i) P(E_j|R_i)\;,
\ee
else the agent could be Dutch-booked as we explained in Section~\ref{DutchieBoy}. This follows because all the lotteries and compound lotteries that define these valuations can in principle be settled.  It is a statement purely about coherence and has nothing to do with physics.

Yet, there is no a priori reason to believe that the agent should assign
\begin{equation}
Q(E_j) = P(E_j)\;.
\label{Gizmoido}
\end{equation}
In the second protocol, there are no compound lotteries by which to even define the numbers $P(E_j|R_i)$.  Thus, if such an identification were to be made, it would require input going beyond probability theory:  It would require {\it physics\/}~\cite{DeBrotaStacey2020,DeBrota2020c}.  In fact, if one considers classical physics in Liouville form and takes the phase-space points of a system to be an analog of our reference apparatus, one arrives at just such an identification.

Eq.\ (\ref{Gizmoido}) then is not a statement of probability theory, but a statement of physics!  Classical physics.\footnote{This is to say, even classical physics can be viewed in a normative light. One is just not accustomed to thinking of it that way.}  One thus suspects that quantum theory, if it gives anything at all beyond the negative statement
\begin{equation}
Q(E_j) \ne P(E_j)\;,
\label{Gizmonico}
\end{equation}
it will give something to replace Eq.\ (\ref{Gizmoido}).

In fact, quantum theory is not at all silent on relating the $Q(E_j)$ to the $P(R_i)$ and the $P(E_j|R_i)$.  Let us exhibit one such relation explicitly.  There are many possibilities depending upon the choice of reference apparatus, but a unique property of them all is that Eq.\ (\ref{Gizmoido}) is {\it never\/} satisfied~\cite{DeBrota2020}.  First consider the second protocol.  Suppose that upon learning the outcome $R_i$ of the first measurement, the agent ascribes a quantum state $\hat\sigma_i$ to the quantum system going forward to the $\hat E_j$ measuring device.    Then, the reference apparatus will be uniquely characterized by an invertible matrix
\be
\label{SuperDuper}
\big[\Phi^{-1}\big]_{ij} := \tr \hat{R}_i \hat{\sigma}_j\;,
\ee
and an elementary calculation reveals that the Born rule, Eq.\ \eqref{SecondBorn}, becomes
\be
\label{ltpanalogindices}
Q(E_j)=\sum_{i=1}^{d^2}\left[\sum_{k=1}^{d^2}[\Phi]_{ik}P(R_k)\right]\! P(E_j|R_i)\;,
\ee
\begin{figure}
\begin{center}
\includegraphics[width=3.5in]{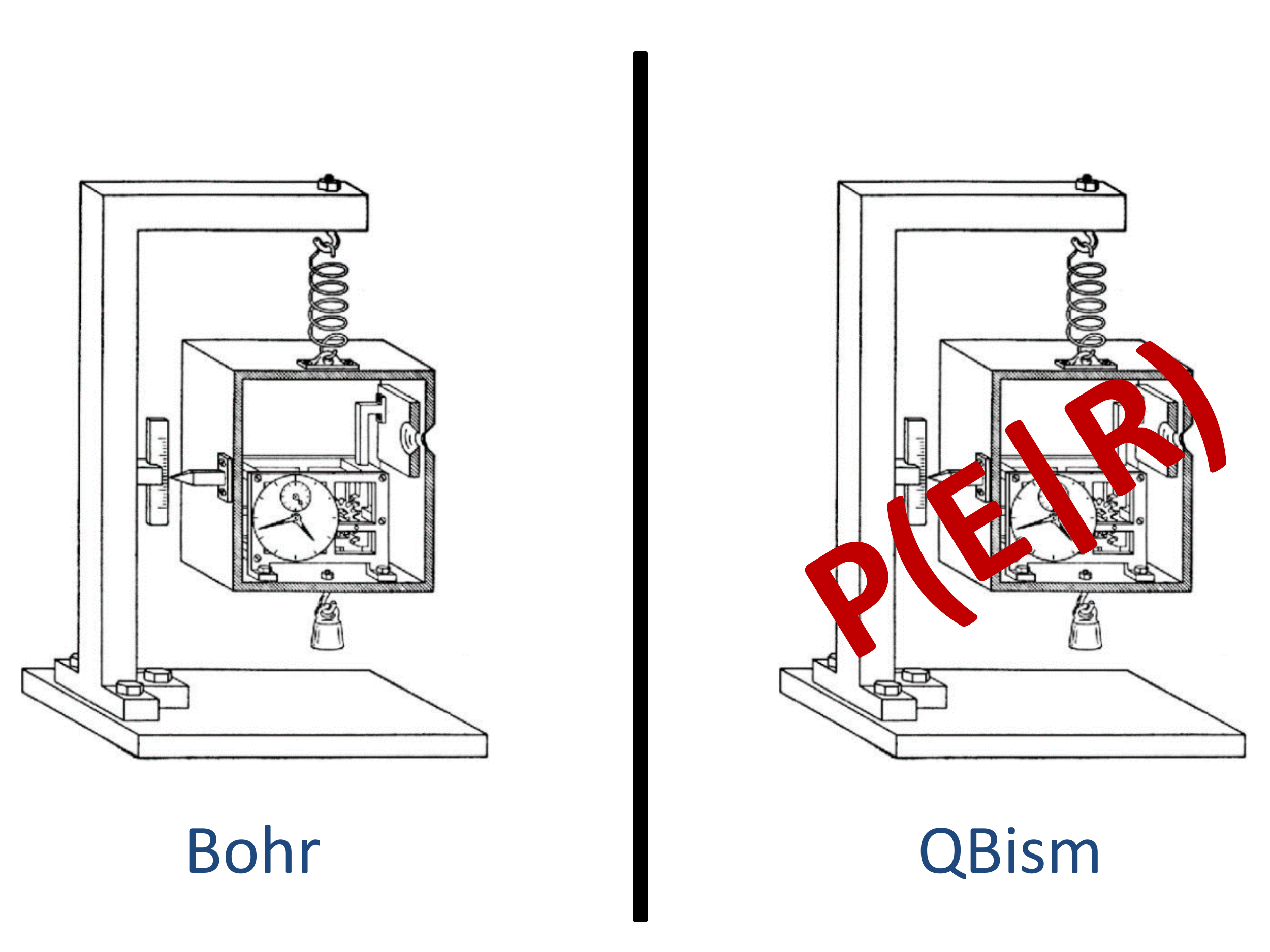}
\end{center}
    \caption[Caption for LOF]{\label{BohrAndQBism} {\bf Skeptic:} ``QBists say that quantum states and measurement operators both represent personal judgments living at the same conceptual level.  But that can't be right!  Perhaps it is true that no one can see a quantum state, but anyone can walk into a laboratory and see a measuring device for exactly what it is.  This is why Bohr~\cite{Bohr49} went to great lengths to depict measuring devices as heavy, bulky instruments, firmly bolted to their laboratory benches.''  {\bf QBist:} ``Bohr was wrong.  Do you know how much implicit and explicit statistical analysis and calibration go into specifying the devices of even a small quantum optics experiment? Who can walk into the lab and see the personal, prior probabilities in a Bayesian experimentalist's head?  The very identification of the device boils down to a very complex set of interlocking probability assignments {\it for him}.\footnotemark\ \ Fortunately there is a formalism that makes the end result explicit: A set of conditional probability assignments $P(E_j|R_i)$, as in Eq.\ \eqref{Lilliputian}, no more or less personal than a quantum state itself.''}
\end{figure}
\footnotetext{In this regard, it can be instructive to look at a ``simple'' quantum information task such as quantum teleportation. First, see it as it is presented on paper~\cite{Teleportation}, where the protocol can be depicted in diagrams almost as simple and stark as Bohr's illustration above. Then compare {\it that\/} to an image of an actual laboratory implementation of the same, such as in this photograph~\cite{TeleportationExperiment} with its more than 500 mirrors and lenses.}

\noindent where
\be
P(E_j|R_i)=\tr \hat\sigma_i \hat E_j\;.
\ee
Moreover, we now know what the judgments $\hat E_j$ are judgments of.  They are just symbolic ways to express the $P(E_j|R_i)$, $i=1,\ldots,d^2$, the agent's conditional lottery valuations defined in the second protocol:
\be
\label{Lilliputian}
\hat E_j \quad\longleftrightarrow \quad  \Big\{P(E_j|R_i)\Big\}_{i=1}^{d^2}\;.
\ee

In all, Eq.\ \eqref{ltpanalogindices} tells us that except for a specification of the details of the reference apparatus, the Born rule is purely a relation between probabilities and conditional probabilities.  Particularly, it means that the content of the Born rule has everything to do with Asher Peres's dictum, ``unperformed experiments have no results''~\cite{Peres78}. Unperformed experiments may not have results, but that does not mean an agent shouldn't be cognizant of how she would gamble if they were to be performed~\cite{Fuchs13a}.  Moreover, see Fig.\ 2 for the immense conceptual change Eq.~(\ref{Lilliputian}) brings to the table of quantum interpretation: It is perhaps the most compelling reason for the QBist contention that a measuring device must be understood as a part of the agent herself.

The form of Eq.\ (\ref{ltpanalogindices}) can be made all the more striking by introducing a completely ``vectorized'' notation, in which omitted subscripts signify an entire vector or matrix. That is, take $[P(E|R)]_{ji}=P(E_j|R_i)$, $[P(R)]_i=P(R_i)$, and $[Q(E)]_j=Q(E_j)$.
Then the Born rule takes the compact form
\be
\label{ltpanalog}
  Q(E) = P(E|R)\Phi P(R)\;.
\ee
Note how this differs from the classical rule in Eq.\ (\ref{Gizmoido}), where the $Q(E)$ in the first protocol is equated with the expression derived from the Law of Total Probability $P(E)$,
\be
\label{Duderwink}
Q(E) = P(E|R)P(R)\;.
\ee

In this language the only difference between the quantum and classical assumptions---they are both normative rules above and beyond Dutch-book coherence---is concentrated in the fact that $\Phi\ne I$ in quantum theory regardless of the reference apparatus.  In fact, there is a minimal finite separation between the two matrices~\cite{DeBrota2020}.\footnote{For those interested see the optional Subsection~\ref{SIC-Business} for more detail.} Yet, therein lies a profound difference between these two possible worlds an agent might inhabit. The distinctive flavor of the quantum world will come into sight when we finally arrive at the Wigner's friend thought experiment in Section~\ref{Wigner's Hands}.

This tenet has led some commentators to claim that QBism is a form of instrumentalism. This, as with the claim of solipsism, is also easily refuted; see, e.g., Refs.~\cite{Fuchs16,Wallace2016}. Indeed as emphasized in the Introduction, from its earliest days the very goal of QBist research has been to distill a statement about the character of the world from the fact that the gambling agents within it should use the quantum formalism~\cite{Fuchs2002}. Why would it be so?  Whatever the answer turns out to be, it will be a statement about the particulars of reality.  If the world were different in character, then the agents within that world would be better advised to use something other than quantum theory for their gambles.

The idea is a simple one for physicists: What QBism aims for is to reverse engineer from the formalism to a characterization of an ontology, while never straying from the progress it has made by viewing quantum theory as an addition to decision theory. This reverse engineering remains an active research program---a sign that QBism is a living subject.

Philosophers of physics seem to have more trouble than physicists with the stark admission that something is an ongoing project, as they seem to desire conclusive answers no matter how ill-considered. ``Tell us your ontology {\it now}, or there is nothing to discuss!,'' as someone like Tim Maudlin would exclaim~\cite{Maudlin2016}. We like to quote Schr\"odinger~\cite{Schroedinger54} as a response:
\bq
\noindent In an honest search for knowledge you quite often have to abide by ignorance for an indefinite period.  Instead of filling a gap by guesswork, genuine science prefers to put up with it; and this, not so much from conscientious scruples about telling lies, as from the consideration that, however irksome the gap may be, its obliteration by a fake removes the urge to seek after a tenable answer.  So efficiently may attention be diverted that the answer is missed even when, by good luck, it comes close at hand.  The steadfastness in standing up to a {\it non liquet}, nay in appreciating it as a stimulus and a signpost to further quest, is a natural and indispensable disposition in the mind of a scientist.
\eq

Nonetheless, the methodology has already led to a number of strong ontological claims on the part of QBism---from the world being capable of genuine novelty and being in constant creation~\cite{Fuchs2017}, to the Born rule expressing a novel form of structural realism.  To put a term on the books and contrast with it the ontic and epistemic varieties of structural realism discussed by the philosophers~\cite{Ladyman2020}, we might call this part of what QBism aims for a {\it normative structural realism}~\cite{Fuchs2021}. As the philosopher Craig Callender once paraphrased the idea, the Born rule would then represent nature's whisper to its agents~\cite{Callender2015}.

\subsubsection{Symmetric Informationally Complete (SIC) Reference Apparatuses}
\label{SIC-Business}

In the previous part of Section~\ref{NormativitySection} we have been completely catholic in our choice of {\it reference apparatus}. Eq.~(\ref{ltpanalog}) holds for of any of them.  We did the general case without further detail so that the bare bones of the extra normativity implied by the Born rule would be on stark display.  That extra normativity comes about in the relevance of the Born rule to Fig.\ \ref{Metzenbaum}, rather than the Law of Total Probability. The root cause of this is that we live in a world where {\it all action matters\/} (Durant)\footnote{See also these two early expressions in the QBist corpus~\cite{Fuchs2002c,Fuchs2007} for a more full-flavored account.}---the reference apparatus as an action is conceptually ineliminable.

However, in much of the QBist literature going back as far as~\cite{Fuchs2009}, extra emphasis has been given to a uniquely interesting class of reference apparatuses {\it should they exist}.  These are the so-called symmetric informationally complete quantum measurements or SICs \cite{Zauner1999,Renes2004,Fuchs2017b} in combination with assuming L\"uders' rule for the post-measurement quantum-state assignments.  Since the exploration of these structures is at the cutting edge of the technical side of QBism and some of the papers in this volume may reference this representation (as for instance Boge's does \cite{Boge2022}), it seems worth giving an explanation of them here. In fact, it is a great laboratory for exhibiting the interplay emphasized in the Introduction between QBism's conceptual development and the mathematical demands it makes upon the theory. The philosophy really can't be divorced from the mathematics.  Finally, this small excursion from the main goals of Section~\ref{LoudScreechingHalt} will give an opportunity to show how the SICs play a conceptual role in some of the latest QBist thinking, where they give rise to a kind of ``QBist Planck's constant'' for finite dimensional quantum systems.

Comparing Eqs.\ (\ref{ltpanalog}) and (\ref{Duderwink}) raises an interesting mathematical question for QBism. Depending
upon which reference device the agent chooses for Eq.~(\ref{ltpanalog}), it
can be made to look less like or more like the classic LTP. If one could find a reference
device so that $\Phi = I$, then one would have the LTP identically, and quantum theory's ``extra'' normative rule would not be extra at all. As already explained, though, there is no such reference apparatus~\cite{Fuchs2002}.

So, how close $\Phi$ can be made to look like the identity matrix $I$? The answer would establish
an important fact about quantum mechanics, namely the ``essential difference''
between the Born Rule and the classical intuition that would seek to set $Q(E) = P(E)$ if it could.

In \cite{DeBrota2020}, this question was quantified by introducing a class of
distance functions based on unitarily invariant norms~\cite{Horn1994},
\be
d(I, \Phi ) = \| I-\Phi \|\, .
\label{Ding-ding-ding}
\ee
A unitarily invariant norm is a matrix norm for square matrices such that $\|UXV\|=\|X\|$ for any unitary matrices $U$ and $V$. Such norms form the most significant class of norms in matrix analysis~\cite{Horn1994}. The class includes the trace norm, the Frobenius norm, the operator norm, all the other Schaaten $p$-norms, and the Ky Fan $k$-norms. The class of $\Phi$ matrices that achieve the minimal distance $d_Q$ from the identity $I$ define the {\it essential quantumness\/} of the Born Rule: It establishes the essential gap between the classical and quantum normative rules.

To set ourselves up for expressing the essential quantumness, let us first define the notion of a SIC\@.  A SIC is a POVM with $d^2$ elements for which all the $\hat R_i$ are rank-1, i.e., of the form
\be
\hat R_i=c_i|\psi_i\rangle\langle \psi_i|\;,
\ee
and for which this stringent symmetry condition holds:
\begin{equation}
\tr \hat R_i \hat R_j = c \quad \forall i\ne j\;.
\end{equation}
When this is so, one can prove that the $\hat R_i$ must be linearly independent, i.e., they form a basis for the space of operators. One can further prove that the value of the $c_i$ are fixed to $c_i=1/d$ and
\be
c=\frac{1}{d^2(d+1)}\;.
\ee
Thus a SIC with post-measurement states $\hat\sigma_i=d \hat R_i$ (which is what would come about if the measurement induces L\"uders' rule in the wake of its action) will make for a perfectly good reference apparatus.

SICs have yet to be proven to exist in all finite dimensions $d$, but they are widely believed to exist without exception~\cite{Fuchs2017b},\footnote{The current state of knowledge is that they have been {\it proven\/} to exist via exact algebraic solutions in over 150 distinct dimensions extending up to $d=39,604$, and there is high precision numeral evidence for all $d\le 193$~\cite{GrasslPrivate}} and have even been experimentally demonstrated in some low dimensions~\cite{Durt2008,Medendorp2011,Zhao2015}.  Proving their existence has been a consternating problem in that the effort has been going on for 23 years now, enmeshing more and more researchers~\cite{Fuchs2017b}.  However, the longer it goes, the more tantalizing the hope becomes that the payoff will be big in terms of previously undreamt of physics. For instance in the last six years a connection between SIC existence and Hilbert's (still open) 12$^{\it th}$ problem---a problem to do with algebraic number theory---has been uncovered and become quite the research rage~\cite{Appleby2017b,Appleby2021}.  This strikes of some very deep mathematics going on here---maybe not Fermat's Last Theorem, but something in that direction---and leaves one with the question of why on earth basic quantum theory should care about exotic algebraic number fields?\footnote{From a different direction, but also to some extent mathematically deep, see~\cite{Appleby2011,Appleby2015}.}  What gem for physical understanding is hidden in this?

Thus we return to the question of the essential quantumness. Let $\Phi_{\rm SIC}$ denote a $\Phi$ for the special case of a SIC reference apparatus.  As you might guess by now, the result of \cite{DeBrota2020} is that for all the distance measures considered in Eq.~(\ref{Ding-ding-ding}) and for all reference apparatuses,
\be
\label{PoppyWood}
d\big(I, \Phi \big)\ge d\big(I, \Phi_{\rm SIC}\big)
\ee
with equality if and only if the reference device measures a SIC and outputs post-measurement states that are also elements of a SIC. Hence, if SICs do in fact exist in all dimensions,
\be
d_Q=d\big(I, \Phi_{\rm SIC}\big)\;.
\ee

In the past, QBism has indeed given special attention to reference apparatuses based on SICs, but in all cases previous to the result of Eq.\ (\ref{PoppyWood}), it was essentially for aesthetic reasons.  For instance, note the particularly simple form Eq.~\eqref{ltpanalogindices} takes in this case:
\begin{equation}
Q(E_j)=\sum_{i=1}^{d^2}\left[(d+1)P(R_i)-\frac{1}{d}\right]\! P(E_j|R_i)\;.
\label{urgleichung}
\end{equation}
Long before the optimization problem was formulated, it was apparent that no representation of the Born rule could be simpler than this or more aesthetically similar to the LTP\@. To a physicist's nose, this was already a worthy-enough lead to follow wherever it might lead~\cite{Fuchs2011a,Fuchs2011b,Appleby2017,DeBrota2021,Stacey2021}.  Now however, to see that this form is not only the solution of a conceptually important optimization problem at the root of QBism, but also appears to be deeply connected to Hilbert's $12^{\it th}$ problem, seems astounding.\footnote{By the way, the simplicity of a form like this arising from the essential quantumness question is not at all a given. A proof of principle can be found in the well-studied foil to standard quantum theory known as real-vector-space quantum mechanics. In that theory, everything in the formalism is identical to standard quantum theory, except that it is based on real vector spaces rather than complex: Vectors always have real components with respect to any basis, symmetric operators take the place of Hermitian operators, and rotations take the place of unitaries. As it turns out, an analogue of a SIC (call it a real-SIC) generally does not exist in that theory. For instance real-SICs do exist for dimensions $d=2,3,7,$ and 23, but can be proven not to exist for any of the dimensions in between. So, what becomes of the essential quantumness in the dimensions for which a real-SIC does not exist?  Recently, we explored the issue for $d=4$ and the result was strikingly ugly~\cite{Fuchs2022}! In that setting, Eq.\ (\ref{urgleichung}) is replaced by an expression requiring four complicated lines of symbols for its display, and even then the expression depends on a floating parameter according to which unitarily invariant norm in Eq.~(\ref{Ding-ding-ding}) one uses. Nothing appears universal about it.  If one had to settle for a representation of the Born rule like that in standard complex quantum theory, one would wonder why even bother?}  Where it will go, no one knows.

\begin{figure}
\begin{center}
\includegraphics[width=5.7in]{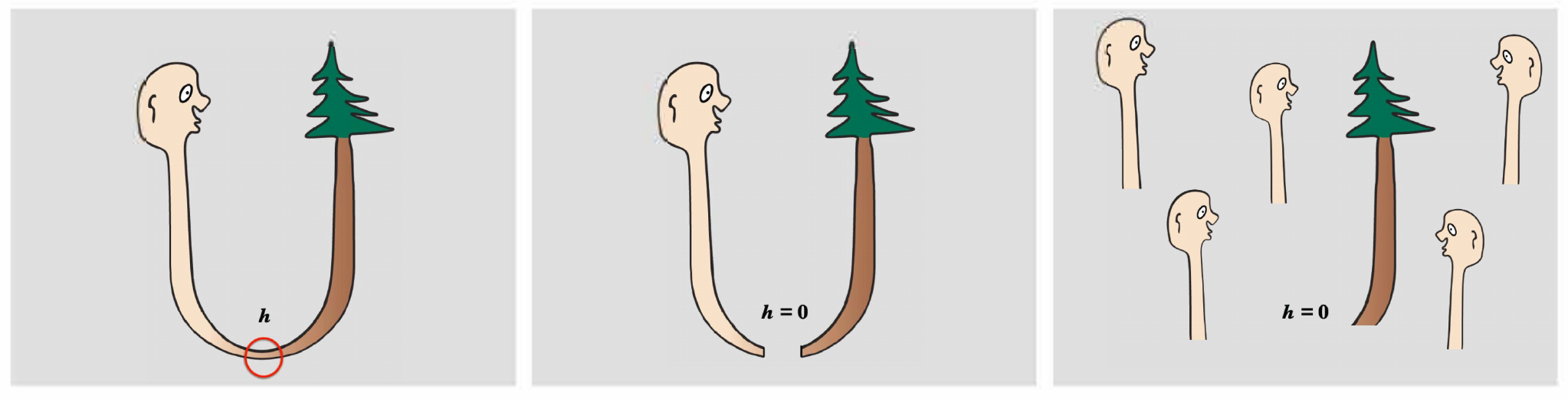}
\end{center}
    \caption{\label{AmandaOnPlanck} Amanda Gefter's parable \cite{Gefter2018} building on John Wheeler's famous U diagram \cite{Wheeler1975,Wheeler1988}. \medskip\\ {\bf Left Frame:} ``You have a subject, and an object, and mostly they seem like separate things, but when you look closely enough you find that they are conjoined, that there's this piece of connective tissue where they can’t be pulled apart because within that region there’s no way to say which is which. \ldots\ The discreteness [we find in QM] is not a property of the object, it's a property of the subject-object relation. I think a useful way of thinking about this is to see Bohr's quantum of action as a kind of coupling constant between subject and object, between observer and observed.'' \medskip\\ {\bf Middle and Right Frames:} ``As you turn the strength of that coupling down, the area of overlap gets smaller and smaller, and if you turn it all the way down to zero, observer and observed can be neatly separated, it recovers classical physics, and it frees the object from its subject so that it can be shared by other subjects.'' \medskip\\ {\bf Moral:} ``But of course $h$ is not zero and so we have to contend with the irreducible coupling between subject and object that Bohr would say is the very essence of quantum physics, and I would say ought to inform our philosophical discussions, in that realism is about an object decoupled from any subject, which you can't do, and idealism is about a subject decoupled from any object, which you also can't do. But what does that leave you?'' \medskip\\ {\bf Analysis:} ``[A]s Wheeler delved deeper into the physics of gravitational collapse, he realized that you can’t avoid singularities, and that at the singularity not only would spacetime itself disappear, but all the laws of physics as we know them---all conservation laws---would disappear, too.  If spacetime is destroyed in gravitational collapse, it wasn't fundamental enough to be the basement-level ingredient of ultimate reality.  And so Wheeler asked himself, when spacetime and all the conservation laws are destroyed in gravitational collapse, what survives?  And the only answer he could come up with was: the quantum principle. What is the quantum principle?  It seemed to go back to the fundamental coupling between subject and object.''}
\end{figure}

In the meantime, we can already get a glimpse of the meaning of $d_Q$ by hijacking and modifying a parable due to Amanda Gefter about Niels Bohr's and John Wheeler's understanding of the ``quantum principle'' \cite{Gefter2018}.  See Fig.\ \ref{AmandaOnPlanck} for Gefter's original story. When I first saw this at a talk in South Africa, my jaw dropped because it so strikingly captured not what Niels Bohr was {\it on about}, but rather what QBism has always {\it been about}.  After all, Bohr always had his ``agencies of observation'' (whose design can be expressed in common language suitably refined by the terminology of classical physics, etc., etc.)\ mediating between the subject and object, but Gefter's diagrams really went for the jugular---they got to the essential point, true subject-object ambiguity, nothing in between.  The metaphor was on the mark.

There remains a question of detail, though, that is worthy of further exploration:  Should the size of the circle in Fig.\ \ref{AmandaOnPlanck} be symbolized by Planck's constant $h$ or perhaps something else? A clue comes from John Wheeler\footnote{Some paraphrasing for punchiness and clarity. Also units have been adjusted to be on the same footing.}~\cite{Wheeler1990}:
\bq
\noindent How come a value for the quantum so small as $\hbar = 1.05 \times 10^{-34} J\cdot s$\@? As well as ask why the speed of light is so
great as $c =3.00 \times 10^6 m/s$\@! No such constant as the speed of light ever makes an appearance in a truly fundamental account of special relativity or Einstein geometrodynamics, and for a simple reason:  Time and space are both tools to measure interval. We only then properly conceive them when we measure them in the same units. The numerical value of the ratio between the second and the meter totally lacks teaching power.  It is an historical accident.  Its occurrence in equations obscured for decades one of nature's great simplicities.
Likewise with $\hbar$\@! One day we will revalue $\hbar = 1.05 \times 10^{-34} J\cdot s$---as we downgrade $c =3.00 \times 10^6 m/s$ today---from constant of nature to artifact of history.
\eq
From this point of view, Planck's constant is just not something that has a meaningful size.\footnote{Perhaps this paper~\cite{Ralston2020} is also relevant?}  Indeed one will notice that neither $h$ nor $\hbar$ made any appearance in this paper until this very discussion.

Yet, Planck's constant surely serves a purpose in {\it applications\/} of quantum mechanics. What's the catch?  For QBism, recall that quantum theory should be thought of as additive to classical theory. Here's the way we put it in~\cite{Fuchs10a}:
\bq
\noindent The expectation of the quantum-to-classical transitionists is that quantum theory is at the bottom of things, and ``the classical world of our experience'' is something to be derived out of it.  QBism says ``No.  Experience is neither classical nor quantum.  Experience is experience with a richness that classical physics of any variety could not remotely grasp.''  Quantum mechanics is something put on top of raw, unreflected experience.  It is additive to it, suggesting wholly new types of experience, while never invalidating the old.  To the question, ``Why has no one ever {\it seen\/} superposition or entanglement in diamond before?,'' the QBist replies:  It is simply because before recent technologies and very controlled conditions, as well as lots of refined analysis and thinking, no one had ever mustered a mesh of beliefs relevant to such a range of interactions (factual and counterfactual) with diamonds.  No one had ever been in a position to adopt the extra normative constraints required by the Born Rule.  For QBism, it is not the emergence of classicality that needs to be explained, but the emergence of our new ways of manipulating, controlling, and interacting with matter that do.
\eq
It's in this that Planck's constant comes into the story for QBism.

Here is an example.  Suppose an agent starts with a system she has been modelling as a classical harmonic oscillator when its energies are not too high (say $< E$). If she has gotten fine-tuned enough with her manipulations of it, she might decide it would be to her benefit to start treating it as a quantum system.  But how should she perform such an ``upgrade?''  First and foremost she will need to associate a Hilbert space with it.  But which one, what dimension?  QBism takes the following as guidance.  Since she's restricting her energies to be less than $E$, the potential trajectories of the system in a phase-space description of it will be bounded by an ellipse appropriate to $E$.  To upgrade to Hilbert space, the agent simply chooses a dimension roughly equal to the area circumscribed by the ellipse divided by $h$, the quantum of action.  It's rough guidance, but it's a start.  One thing it emphasizes though is that $h$ is discarded as soon as one starts to consistently invoke quantum theory.

So, in a way, it is Hilbert space itself that represents Gefter's circle, signifying the ineliminable connective tissue that is neither subject nor object.  But we can do still better with the tools developed in this section at the same time as remaining faithful to the original metaphor.  Even with a finite-dimensional Hilbert space established, the Born rule has a range of expression for how much the agent {\it appears to be\/} ``coupled'' (Gefter's term) to the system during the process of measurement.  But one might say those are more ``coordinate effects'' than anything else. The role of expression (\ref{urgleichung}) is to establish the genuine ``strength'' of the coupling.  It is not a single number like $\hbar$ that we are dealing with now, but it still expresses some way to quantify a contingent feature of our world:  Namely, just how much subject and object can be individuated from their substrata.  This is the kind of contingent feature of nature calling out from quantum theory that the various philosophical ontologies courted in our concluding Section~\ref{DuhDuhDuh} could not have known about.  The question now is how we might incorporate these details into the insights of the various ``experience first'' philosophers and end up with something even better?

\subsection{Unitary Evolution Expresses an Agent's Degrees of Belief}
\label{SuperBowlSunday}

In the years of slow progress that led to QBism, a number of arguments were harnessed to compel the idea that quantum states should be
understood epistemically---it wasn't just a bald assertion made without any history.  For instance, the quantum no-cloning
theorem~\cite{Caves1996} and the existence of quantum teleportation~\cite{Fuchs2001} were two favorites for making the case in the early days.
But if so, what was to be made of an analogous theorem in quantum information theory that unknown unitary operators could not be
cloned~\cite{DAriano01,Chiribella08}, just as quantum states could not?  Or of the protocols that showed that unknown unitary operators could themselves be teleported~\cite{Huelga01}, again just as quantum states?  QBism's answer was, ``Go it, and go it stronger!''\footnote{As William James said he almost cried aloud with ``glee'' and ``admiration'' during the 1906 San Francisco earthquake~\cite[p.\ 3]{Richardson2006}.}  There was no choice~\cite{Fuchs2002} but to accept that unitaries too---like the quantum measurement operators before them---were cut from the same
cloth as quantum states: All three entities represented personal judgments, not properties of the agent's external world.  Moreover, they all
must reside at the same conceptual level in the agent's attempt to conform to the normative edict of the Born rule.

As before, this is most easily seen through the formalism developed in Section~\ref{NormativitySection}.  Suppose an agent assigns a quantum state $\hat\rho$ as the glue for her probability assignments to the outcomes of measurements she might perform on a system at time $t_0$. If the agent believes that at a later time $t_1$ she won't have gained or lost any predictability on balance for those same measurements, then the time lapsed must correspond to a transformation drawn from the symmetry group of the convex set of quantum states~\cite{Appleby2017}---i.e., the projective unitary group.  That is, at time $t_1$ she should assign a quantum state $\hat U\hat\rho\,\hat U^\dagger$ instead, for some unitary operator $\hat U$.

To put this statement into normative terms, it means assigning a probability distribution
\be
P_{t_0}(R_i) = \tr \hat\rho\hat{R}_i
\ee
to the outcomes of the reference measurement at $t_0$ and another other probability distribution---let us call it $P_{t_1}(R_i)$---at time $t_1$,
\be
P_{t_1}(R_j) = \tr \big(\hat U\hat\rho\,\hat U^\dagger\big)\hat{R}_j = \tr \hat\rho\,\big(\hat U^\dagger\hat{R}_j \hat U\big)\;.
\ee
This suggests that we focus on the measurement $\hat{R}^\prime_j = U^\dagger\hat{R}_j\hat U$ and think of it just as any of the measurements we could put into the right side of Fig.\ \ref{Metzenbaum} to get:
\be
P_{t_1}(R_j)=Q(R^\prime_j)\;.
\ee
With this it becomes obvious what the relation between the $P_{t_0}(R_i)$ and $P_{t_1}(R_j)$ must be.  It is just the Born rule as expressed in the language of Eq.\ \eqref{ltpanalogindices}, but with slightly modified variables:
\be
\label{UnitarityThisTime}
P_{t_1}(R_j)=\sum_{i=1}^{d^2}\left[\sum_{k=1}^{d^2}[\Phi]_{ik}P_{t_0}(R_k)\right]\! P(R^\prime_j|R_i)\;,
\ee
where
\be
\label{RiffRaff}
P(R_j^\prime|R_i)=\tr \hat\sigma_i \hat{R}_j^\prime\;,
\ee
and $\Phi^{-1}$ is defined just as before in Eq.~(\ref{SuperDuper}).  In the special case when a SIC is used as the reference apparatus, this becomes
\be
P_{t_1}(R_j)=\sum_{i=1}^{d^2}\left[(d+1)P_{t_0}(R_i)-\frac{1}{d}\right]\! P(R^\prime_j|R_i)\;.
\label{timetrip}
\ee

With this, {\it the very meaning of unitary evolution\/} is at hand: It is captured all and only by the conditional probability assignments in Eq.\ \eqref{RiffRaff}.  That is, it is transparently as much a personal judgment as the quantum states assignments $P_{t_0}(R_i)$ and $P_{t_1}(R_j)$ are.  Moreover, the integrated Schr\"odinger equation $\hat\rho\,\rightarrow\,\hat U\hat\rho\,\hat U^\dagger$ in this notation is nothing other than a special case of the Born rule itself:
\be
\label{UnitaryBoy}
  Q(R^\prime) = P(R^\prime|R)\Phi P(R)\;.
\ee
Via similar considerations, one can also show that an equation of exactly the same form holds for the full class of all possible quantum time evolutions, not just unitary evolutions. In the quantum information literature, these are known as completely positive trace-preserving maps or {\it quantum operations}. As it must follow because of Eq.~(\ref{UnitaryBoy}), all such maps are to be understood as subjective judgments within QBism.  In fact, just as proving a de Finetti representation theorem for ``unknown quantum states'' was a crucial test for QBism in its earliest days, one can prove an analogous result for ``unknown quantum operations''~\cite{Fuchs2004a,Fuchs2004b} thereby ensuring the consistency of the point of view.

As a final point, let us note a major distinction between the QBist way of thinking about unitarity and the workaday quantum information scientist's way.  In the latter venue a unitary operation is something that is engineered, for instance via the design of a quantum circuit.  One then builds a physical implementation of the circuit in the laboratory to achieve some task or to ``control'' an unwieldy system.  Speaking of it in that manner, it may be hard to stop employing (even implicitly) an ontic understanding of unitary operations: They are expressions of the {\it solid stuff\/} that makes things happen.

However, the word ``control'' makes no appearance in the QBist account of unitarity. In QBism, the assignment of a unitary operation is just that:  An {\it assignment}, a judgment that an agent makes to reasonably represent her beliefs. It is not an {\it action\/} she takes upon a system, forcing it to do one thing or another, as is the case with a quantum measurement (with its generally unpredictable outcome).  This stance also sets QBism apart in the spectrum of quantum interpretations. To anticipate one aspect of the Wigner's friend thought experiment in Section~\ref{Wigner's Hands}, a QBist for instance would never say, as Baumann and Brukner do in~\cite{Baumann2020},
\begin{quote}
We note that the specific relative phase between the two amplitudes \ldots\ is determined by the interaction Hamiltonian between the friend and the system, which models the measurement and is assumed to be known to and {\it in [the] control of\/} Wigner. [our emphasis]
\end{quote}
Wigner is in as much control of his external world with a unitary-operation assignment as he is with any other personal probability assignment---namely, none.  A unitary instead represents an agent's hard won degrees of belief. Those beliefs may make the agent very certain of the outcome of an appropriate measurement, but a belief can always be shattered by a surprise.

\subsection{Even Probability-One Assignments Are Judgments}
\label{EveningBreeze}

No Bayesian of any of the 46,656 varieties classified by I.~J. Good~\cite{Good1983} would disagree that a probability assignment $P(D)=p$ for some $p$ in the range $0<p<1$ represents a personal judgment---i.e., that it is not a fact of nature.  Things only become testy when one starts to discuss the limit points 0 and 1, and that is where QBism makes one of its most distinctive stands.  QBism regards even probability 0 and 1 assignments as personal judgments.

Assigning probability-1 to an outcome expresses the agent's supreme confidence that the outcome will occur, but it does not imply that there is something in nature to guarantee that the outcome {\it actually\/} must be the case.  QBism makes a strict category distinction between the truth (or facticity) of measurement outcomes and probability-1 (as a belief) that other Bayesian interpretations, say that of Jaynes~\cite{Jaynes2003}, ignore.

It was first recognized that such an extreme interpretation for probability-1 was crucial for QBism's consistency in 2001~\cite{Fuchs2002b}, and later it turned into an indispensable tool for analyzing the locality of quantum theory~\cite{Fuchs14a}, but QBists have only recently learned that this notion already had a respectable pedigree in 1991 in the thinking of the philosopher and logician Richard Jeffrey~\cite{Jeffrey1991}:
\bq
\noindent
The (``Bayesian'') framework explored [here] replaces the two Cartesian options, affirmation and denial, by a continuum of judgmental probabilities in the interval from 0 to 1, endpoints included---or what comes to the same thing---a continuum of judgmental odds in the interval from 0 to $\infty$, endpoints included.  Zero and 1 are probabilities no less than 1/2 and 99/100 are.  Probability 1 corresponds to infinite odds, 1:0.  That's a reason for thinking in terms of odds:  to remember how momentous it may be to assign probability 1 to a hypothesis.  It means you'd stake your all on its truth, if it's the sort of hypothesis you can stake things on.  To assign 100\% probability to success of an undertaking is to think it advantageous to stake your life upon it in exchange for any petty benefit.  We forget that when we imagine that we'd assign probability 1 to whatever we'd simply state as true.
\eq

Mundane examples of the distinction between probability-1 and truth abound so long as one is not wedded to an objective, agent-independent notion of probability.  Here is an example from the world of prenuptial agreements.  Anyone entering a marriage knows that their partner, being another free agent, is free to be faithful or cheat in the relationship.  Marriage is a decision one does not make lightly. Thus, consider an agent who through an extensive exploration of the largest set of intertwined beliefs she can muster---all the many things she believes of her partner, the things she believes of her partner's family, their religious views, known financial matters, the society in which they live, \ldots\ any number of things, even weighing the advice of her lawyer---aims to make a probability assignment for her partner's being faithful.  It would be an enormous computational task, but never mind that.  What is clear from personalist Bayesian probability theory is that when a mesh of beliefs is wide enough, it can be enormously restrictive to one's probability assignments~\cite{Lindley2006}.  In this case, imagine our agent ends up with $p=1$ that her partner will stay faithful.  Yet, the agent's lawyer advises a prenuptial agreement.  Is the agent wrong?

What does this have to do with physics?  Because of the intimate connection between quantum states and the probabilities derived from them, QBism regards the assignment of all quantum states (mixed or pure) as an agent's personal judgments. This implies in particular that even a statement such as ``this outcome is certain to occur for this quantum measurement'' reflects an agent's judgment rather than a fact of nature. In other words, nothing in nature metaphysically guarantees that an outcome to which an agent has assigned probability-1 will in fact occur.  As the world is genuinely indeterministic according to quantum theory, an agent's judgments are genuinely fallible.

This point cannot be emphasized enough, as it seems a number of commentators on QBism simply do not recognize its importance for the QBist
response to a number of conundrums.
For instance, consider an agent who takes the action of placing a Stern-Gerlach device in front of an electron and has just
registered spin-up for it in the $z$-direction as her consequent experience.  She will thus assign a quantum state $|z=+1\rangle$ for any
subsequent measurements on the electron.  However, in QBism this does not amount to a statement of fact but a statement of belief.  The
assignment of this state amounts to, among other things, a belief---a monumentally strong belief---that taking the same action with the
Stern-Gerlach device will give rise to exactly the same consequence, namely the experience of spin-up in the $z$-direction.  But what means
``same?''  Even ``same'' is a judgment if one is going to have a consistent subjective view of quantum states~\cite{Fuchs2004a}.  Thus QBism
must say that the notion of ``same measurement'' is itself a belief, not a fact of nature~\cite{Fuchs2002}.  It might be a supremely strong belief
because a long measure-remeasure sequence has given the same result an inordinate number of times previously, but from the QBist conception
this does not negate that it is a belief.  Think, for instance, of Hume's argument that induction can never be guaranteed by any principle other than itself~\cite{Mermin2014}.

Thus, the world from a QBist conception can always surprise, no matter the certainty of the agent using the quantum formalism---and this even if the formalism is applied to something vastly more complicated than an electron.  Writing down $|z=+1\rangle$ does not ensure that an electron will deliver at the agent's command if he performs the right measurement, but only with what the agent believes with his heart of hearts.  Why should it be different for an atom, for a molecule, for a long piece of DNA, for a euglena, or even Wigner's friend?  If we grant autonomy to a single electron so that it might genuinely surprise an agent, why would we not for a monstrously complicated system like Wigner's friend?  One of the lessons of this paper will be that whatever QBism says of an electron, it must say the same of Wigner's friend, and reciprocally whatever we learn from Wigner's friend we should take to our understanding of the world at large.

\subsection{Subjective Certainty of What an Outcome Will Be Does Not Negate that Unperformed Measurements Have No Outcomes}
\label{OdeToAsher}

Asher Peres's slogan, ``unperformed experiments have no results''~\cite{Peres78}, had a powerful influence on the development of QBism.  Maybe in the language of the present paper we should say ``unperformed measurements have no outcomes'' instead, but the sentiment is obvious.  Here is the nice way David Mermin put it in our joint paper~\cite{Fuchs14a}:
\begin{quote}
{\it ``This experiment has no outcome until I experience one.''}
QBism personalizes the famous dictum of Asher Peres. The outcome of an experiment is the experience it elicits in an agent. If an agent experiences no outcome, then for that agent there is no outcome. Experiments are not floating in the void, independent of human agency. They are actions taken by an agent to elicit an outcome. And an outcome does not become an outcome until it is experienced by the agent. That experience {\it is\/} the outcome.
\end{quote}

This has all been said in previous sections, but it does not hurt to explore it from every angle.  What has not been discussed so far is how this conception interacts with the notion of probability-1.  Is there any special difficulty introduced when one considers such extreme probability assignments?  It is common enough to find such a sentiment in the literature.  An example contemporaneous with when QBism first came to its realization is this quote by Brukner and Zeilinger~\cite{Brukner01}, whose positions in other aspects sometimes come somewhat close to QBism:
\begin{quote}
Only in the exceptional case of the qubit in an eigenstate of
the measurement apparatus the bit value observed reveals a
property already carried by the qubit.  Yet in general the value
obtained by the measurement has an element of irreducible
randomness and therefore cannot be assumed to reveal the bit
value or even a hidden property of the system existing before
the measurement is performed.
\end{quote}
Maybe they have developed since then, but historical examples can be found in the writings of Heisenberg, Dirac, von Neumann, Messiah, and many others~\cite{Gilton2016}.  It is what the philosophers of physics call the ``eigenstate-eigenvalue link.''

QBism predictably takes the stand that even when an agent assigns probability-1 to one of the possible outcomes of a measurement, there is
nothing in the agent's external world that metaphysically ensures it to come about.  For ``unperformed measurements have no
outcomes'' is a statement about the character of the world---that it is not a block universe---whereas a probability-1 assignment is only a {\it belief\/} (supremely strong, but nonetheless a belief) someone happens to have in the moment.  To say it differently, one is an expression about the world's creative character, while the other is about a user of quantum theory's momentary state of mind.

Of course, the latter expression comes from our analysis in the previous subsection of probability-1 in general.  But the former goes much deeper: QBism in fact takes it to be the great lesson of all the multitude Bell-inequality and Kochen-Specker analyses and experiments of the last half century.\footnote{Perhaps the essential point is expressed the most vividly in this paper by N. D. Mermin~\cite{Mermin81}.} More contemporaneously, QBism sees this lesson further reinforced by the recent ``no-go theorems'' of Pusey, Barrett, and Rudolph (PBR)~\cite{PBR12} and Colbeck and Renner (CR)~\cite{Colbeck2012}, perhaps to the dismay of the authors' original intentions.\footnote{From our point of view what these two no-go theorems prove is that one cannot have the {\it conjunction\/} of two statements: A) that quantum states are epistemic, and B) that
they are {\it epistemic about\/} (i.e., knowledge, information, or beliefs about)  some pre-existent ontic variables.  Holding fast to the notion
that quantum states must be epistemic, it then follows that they cannot be epistemic about pre-existent ontic variables.  Put another way,
these theorems give us reason to reject the ``ontic models framework'' of Harrigan and Spekkens~\cite{Harrigan2010} where unperformed experiments {\it do have results}, not reject that quantum states are epistemic. It is noteworthy that the original titles for the original postings of PBR and CR were ``\href{https://arxiv.org/pdf/1111.3328v1.pdf}{The quantum state cannot be interpreted statistically}'' and ``\href{https://arxiv.org/pdf/1111.6597v1.pdf}{Completeness of quantum theory implies that wave functions are physical properties},'' respectively. Neither title could be sustained in the light of QBism.} Yet, QBism traces these results back to the more primordial idea that the normative advice of quantum theory is given by Eq.\ \eqref{ltpanalog}, instead of the classical advice Eq.\ \eqref{Duderwink} co-opted from the Law of Total Probability (LTP).

Eq.\ \eqref{ltpanalog} used the tools of quantum information theory to express the Born rule as a relation between probabilities that works for any quantum state and any possible measurement, but the idea that the inequivalence of the Born rule to the LTP is the root cause of {\it all the quantum mysteries\/} is an idea that goes back at least to Richard Feynman in the 1940s.  In a 1951 paper titled ``The Concept of Probability in Quantum Mechanics,'' Feynman writes~\cite{Feynman51},
\bq
The new theory asserts that there are experiments for which the exact outcome is fundamentally unpredictable, and that in these cases one has to  be satisfied with computing probabilities of various outcomes.  But far more fundamental was the discovery that in nature the laws of combining probabilities were not those of the classical probability theory of Laplace.

I should say, that in spite of the implication of the title of this talk the concept of probability is not altered in quantum mechanics.  When I say the probability of a certain outcome of an experiment is $p$, I mean the conventional thing \ldots.  I will not be at all concerned with analyzing or defining this concept in more detail, for no departure from the concept used in classical statistics is required.

What is changed, and changed radically, is the method of calculating probabilities.
\eq

Of course, Feynman is expressing the transition to the amplitude calculus here.  In his original 1948 paper on path integrals~\cite{Feynman48}, he describes it as ``essentially a third formulation of non-relativistic quantum theory'' (after matrix and wave mechanics).  What is essential to us is the way he directly contrasts the new combination laws to the LTP\@.  He does this by considering three successive experiments $A$, $B$, $C$, with outcomes $a$, $b$, $c$, and denotes the conditional probability for finding $b$ given $a$ as $P_{ab}$, etc.
In a prescient passage, Feynman~\cite{Feynman48} writes,
\bq
Now, the essential difference between classical and quantum physics lies in [the LTP]\@.
In classical mechanics it is always true. In quantum mechanics it is often false. We shall denote the quantum-mechanical probability that a measurement of $C$ results in $c$ when it follows a measurement of $A$ giving $a$ by $P_{ac}^q$. [The LTP] is replaced by this remarkable law: There exist complex numbers $\varphi_{ab}$, $\varphi_{bc}$, $\varphi_{ac}$ such that
\be
P_{ab}=|\phi_{ab}|^2\,, \quad P_{bc}=|\phi_{bc}|^2\,,\quad\mbox{and}\quad  P_{ac}^q=|\phi_{ac}|^2\;.
\ee
The classical law \ldots
\be
\label{bbb}
P_{ac}=\sum_b P_{ab} P_{bc}
\ee
is replaced by
\be
\label{aaa}
\varphi_{ac}=\sum_b \varphi_{ab} \varphi_{bc}\;.
\ee
If \eqref{aaa} is correct, ordinarily \eqref{bbb} is incorrect. The logical error made in deducing \eqref{bbb} consisted, of
course, in assuming that to get from $a$ to $c$ the system had to go through a condition such that $B$ had to have some definite value, $b$.
\ldots

Looking at probability from a frequency point of view \eqref{bbb} simply results from the statement that each experiment giving $a$ and $c$, $B$ had some value.\footnote{Of course, QBism here relies on a coherence argument instead of a frequentist one, but the point is the same.}  The only way \eqref{bbb} could be wrong is the statement, ``$B$ had some value,'' must sometimes be meaningless. Noting that \eqref{aaa} replaces \eqref{bbb} only under the circumstance that we make no attempt to measure $B$, we are led to say that the statement, ``$B$
had some value,'' may be meaningless whenever we make no attempt to measure $B$.
\eq

In Feynman's case, as in ours, the lesson is the same:  If one adopts any method of gluing one's hypothetical gambles together which does not boil down to an application of the Born rule, then one does it at one's own peril. Any probability assignment for a potential experience, including $Q(E_j)=1$, had better come about by some association $R_i\,\leftrightarrow\, \big(\hat R_i,\hat\sigma_i\big)$ and $E_j\,\leftrightarrow\,\hat E_j$ along with an application of Eq.\ \eqref{ltpanalogindices}, or one is violating the Born rule and consequently ignoring quantum theory's normative advice.

Even when an agent is certain what the outcome will be, unperformed measurements have no outcomes.

\subsection{Quantum Theory Is a Single-User Theory for Each of Us}
\label{SingleUser}

It hardly needs to be said by now, but every formulation we have made in the previous tenets has always referred to the concerns of a single agent.  Even when we briefly considered a ``team of scientists sharing notebooks'' in Section \ref{KikiMowing}, we treated the collective as a single agent.  This is an inheritance from QBism's insistence that quantum theory should be understood as a normative addition to personalist Bayesian probability theory.  D. V. Lindley, a prominent Bayesian statistician, put the point succinctly~\cite{Lindley82},
\begin{quote}
The Bayesian, subjectivist, or coherent, paradigm is egocentric. It is a tale of one person contemplating the world
and not wishing to be stupid (technically, incoherent). He realizes that to do this his statements of uncertainty must
be probabilistic. This is important on its own for it rules out a large class of behavior patterns, like sampling-theory statistics, but is it enough? Once I coined the aphorism ``Coherence is all.'' Was I right?
\end{quote}
QBism clearly leans in that direction, modulo the addition of the Born rule.  But, Lindley goes on to say,
\begin{quote}
It is when we consider two coherent Bayesians that new features arise. Does their egocentric behavior allow them to talk to one another?
In one respect it does. Suppose that you and I are both coherent and you tell me that your probability for $A$, were $B$ to be true, is $\alpha$, say. How does this knowledge affect my probability for $A$ given $B$? It is easy for me to do the coherent calculations in terms of my assessments [i.e., my probabilities] that were $A$ and $B$ both true, you would say $\alpha$, and that were $\neg A$ and $B$ both true, you would say $\alpha$. For to me $\alpha$ is just data and the likelihood ratio updates my probability in the usual way. A weather forecaster who announces rain whenever it is subsequently dry, and vice versa, is badly calibrated but very useful. So I can respect your egocentricity; and you, mine.
\end{quote}
With this, QBism agrees wholeheartedly.  Indeed, one may view the recent QBist analyses~\cite{DeBrota2020b} of the Baumann-Brukner~\cite{Baumann2020} and Frauchiger-Renner~\cite{Frauchiger2018} variations on Wigner's friend as extended exercises in seeing to it that Wigner and his friend respect each other's egocentricities.  The friend may ask Wigner his quantum-state assignment for this or that, but if she does so it can only come about by her taking a physical action upon him (as with
any measurement) and using the subsequently induced outcome as a datum that {\it may\/} or {\it may not\/} make much impact on her
future quantum-state and probability assignments.  {\sl The extent to which it will\/} will depend upon the details of her larger mesh of beliefs, as it does in Lindley's example.  It follows that there can be no overarching principle or prescription for how two agents' quantum states for the same system must be related to each other, and QBism takes this fully onboard.  Quantum theory is a single-user theory for each of us.

Contrast this attitude of QBism's to some earlier---what one might call ``halfhearted''---approaches to understanding quantum states in epistemic terms.  This line of thinking was initiated by Rudolf Peierls, who wrote, ``there are limitations to the extent to which [two agents'] knowledge may differ''~\cite{Peierls91a}:
\begin{quote}
It is possible for two observers to have some knowledge of the same system, and the knowledge possessed by one may differ from the other. \ldots\ In this situation the two observers will use different density matrices. \ldots\ However, the information possessed by the two observers must not contradict the uncertainty principle. \ldots\ This limitation can be expressed concisely by saying that the density matrices appropriate to the two observers must commute with each other. \ldots\ At the same time, the two observers should not contradict each other.  This means the product of the two density matrices should not be zero.~\cite{Peierls91b}
\end{quote}
By simple examples from quantum cryptography~\cite{Mermin01}, one can see that the first part of the Peierls ``compatibility
criterion''---i.e., the part about necessary commutivity---cannot be maintained.  But that didn't stop a number of authors from thinking that
even if Peierls wasn't quite right, there should still be some criterion from physics alone to limit how much quantum-state assignments could
differ.  For instance Brun, Finkelstein, and Mermin (BFM) in~\cite{BFM02} write, ``We derive necessary and sufficient conditions for a
group of density matrices to characterize what different people may know about one and the same physical system,'' and ``[A] special case of
our condition is that two pure-state density matrices are compatible if and only if they are identical.''  In that work, there was an implicit
accommodation to an Everettian-style world-view (anathema to QBism to begin with), but a further analysis~\cite{CFS02,Stacey2016} pointed
out that the BFM-compatibility criterion could in fact be given a non-Everettian, purely Bayesian interpretation, so long as one recognized
that it was only one among a number of inequivalent criteria for two agents to judge when they are in sufficient agreement.  It followed that there was nothing necessary and sufficient about any of the criteria as expressions of ``physics alone.''

Ultimately, by choosing to completely sever any connection between quantum-state assignments and external, agent-independent facts, QBism had to reject the idea of compatibility requirements for distinct agents' state assignments.  The two had to go hand in hand.  In QBism, an agent is judged as making a proper use of quantum theory not by which quantum states she assigns, but by whether she consistently makes use of the formalism when making those assignments.  What her neighbor does in making his assignments is immaterial to her.

To add a sidelight to this, let us give an example~\cite{Fuchs13a} for anyone who still holds out hope that what we have argued is just so much sophistry: One can see that the various compatibility criteria were always unstable notions to begin with.  Consider two agents sitting in a common
laboratory, and in agreement on everything to do with a qubit measurement apparatus in front of them---they each make the same POVM assignment
to it. It is thus within each agents' rights to consider the measuring apparatus an extension of him or herself.  Suppose it to be a
measurement of the $\{|0\rangle,|1\rangle\}$ basis.

Nonetheless, imagine the agents do differ on the quantum-state assignments they make for a two-qubit
system, one qubit of which they are going to measure, while keeping the other safely in the distance.  Say one agent professes a quantum state $\rho_+$ while the other $\rho_-$, where
\be
\rho_\pm=\frac12\Big(|0\rangle\langle0|\otimes|0\rangle\langle0|\,+\,|\pm\rangle\langle\pm|\otimes|\pm\rangle\langle\pm|\Big)
\ee
and
\be
|\pm\rangle=\sqrt{\frac{1}{2}}\big(|0\rangle\pm|1\rangle\big)\;.
\ee
It is easy to check that these two states are compatible by all the criteria considered in~\cite{CFS02}---the BFM criterion, the post-Peierls criterion\footnote{Post-Peierls compatibility is the one later popularized by Pusey, Barrett, and Rudolph~\cite{PBR12} in their attempt to give a no-go theorem for epistemic interpretations of quantum states.}, and others.

Let us consider the case where the first qubit is measured and outcome 1 is found---both agents deemed 0 and 1 possible, but 1 was in fact found.  Then the two agents' post-measurement states for the second qubit (the one not touched yet) will be $|+\rangle\langle+|$ and $|-\rangle\langle-|$, respectively.  With the outcome of a single measurement, two agents go from claiming to having compatible information by every criterion, to being incompatible by all of them!\footnote{Except, that is, for the uniquely QBist criteria $W$ and $W^\prime$ in~\cite{CFS02}, where any two quantum states whatsoever are compatible.}  Of course, this is simply because $|+\rangle\langle+|$ and $|-\rangle\langle-|$ are orthogonal to each other, and for a subsequent measurement of the $\{|+\rangle, |-\rangle\}$ each agent will assign probability-1 for an outcome that the other agent assigns probability-0.

What does a QBist make of this?  Nothing more than has already been said:  We have two agents who have made ``momentous'' probability assignments (in the words of Richard Jeffrey from Section~\ref{EveningBreeze}).  This means that the agents are willing in principle to stake their lives on seeing what they expect to see, but it does not make one of the agents ``right'' and the other ``wrong'' before any measurement is performed.  Nor will it make one ``right'' and the other ``wrong'' after the measurement, unless one equates right and wrong with living and dying.  What else can one expect in a genuinely indeterministic world?  So long as each agent had a coherent mesh of beliefs leading up to their original quantum-state assignment, then they have done all that they could do, and there a QBist must leave it.  In a number of recent ``no-go'' theorems, part of the drama as it is presented is that one agent will predict something with probability-1, whereas another agent will predict it with probability-0, and this is taken to be an affront to quantum theory.  But such things are a matter of course in QBism.

To recap:  All that each individual agent has to go on in QBism is whether they have been as coherent with all applications of probability theory and the Born rule as they can be.  For QBism, quantum theory is always a single-user theory, but it is so for each of us.  Wigner should strive to be coherent in all his own beliefs; his friend should strive to be coherent in all hers. But there is no principle of physics that requires Wigner and the friend to have compatible quantum-state assignments:  Physics suffers none by agents' incompatibilities.

\section{QBism's Eightfold Path and the Missing Noble Truth}
\label{Intermezzo}

To summarize Section~\ref{LoudScreechingHalt}, we have identified eight key tenets of QBism (in a different order this time around):
\begin{enumerate}
\item
A quantum measurement is an agent's action upon its external world.  A quantum measurement is cut from the same cloth as any other action she might take upon her world, as for instance by crossing a street.  What makes an action specifically ``quantum'' is when it is worth the agent's while to analyse her expectations for its outcomes in terms of the quantum formalism.
\item
A measurement's outcome {\it just is\/} the consequent personal experience of the agent taking the action.  In QBism there is no notion of a mediating device between the agent and the quantum system (``decohering'' because of an interaction with an environment). Measuring devices are considered conceptually as parts or extensions of the agent.
\item
A quantum state is an agent's personal judgment. It serves to tie together all her probability assignments for the outcomes of all measurement actions she might take upon a system.
\item
Even unitary evolution operators (or more generally completely positive trace-preserving maps) are agents' personal judgments.  They are not set by nature; they are not ontic. In any individual case, they are set by the agent's expectations across time.
\item
The quantum formalism is normative rather than descriptive.  It guides the agent toward making sure all her judgments are consistent with one another and with the ``quantum'' nature of the world.  This is most easily seen by rewriting the quantum formalism in a form that is purely probabilistic in nature, without operators on complex vector spaces, etc. Quantum states thus become probability distributions for the outcomes of a reference apparatus, measurement operators become conditional probabilities with respect to the same, and the Born rule becomes something clearly seen as a variation on the classic Law of Total Probability.  Furthermore, rewriting the formalism this way nods to future physical enquiry by giving a way of quantifying the essential ``quantumness'' conveyed by the Born rule.
\item
Even probability-1 assignments are judgments and thus metaphysically fallible. Probabilities are so disconnected from the world, they can never tell nature what to do. Probability-1 refers all and only to a momentous judgment for the agent who makes it; it means she will gamble her all on it.
\item
Even when an agent is subjectively certain what the outcome of her measurement action will be, she still may not assume that the yet-to-be-performed measurement has a pre-existent determined outcome.  This is part of the content of using the Born rule in place of the classic Law of Total Probability.
\item
Quantum theory is a single-user theory for each of us.  My use of quantum theory concerns my expectations for my personal experiences.  Your use of quantum theory concerns your expectations for your personal experiences. There is no requirement that those two uses must somehow dovetail together.
\end{enumerate}

It was quite accidental that the number of tenets turned out to be eight, but so be it.  Let us use this as a symbol.  Maybe it's a poor joke, but the first thing that comes to mind is the Buddha's eightfold path!  It is not that QBism is already ``blue pure perfect'' by any means, but this much is established:  The historical reason for the eight tenets as they accumulated was in every case to ``relieve the suffering'' caused by one or another quantum conundrum.

However, the eightfold path in Buddhist doctrine was always subsidiary to the four noble truths. The eightfold path was a methodology for living in the world, but the noble truths were a statement about how the world is. It is to this project that we turn now.  What is QBism's long-sought-for noble truth?  In the past we have expressed most of our ontological thinking in the language of American pragmatism and radical empiricism.  This time, however, we will start with Wigner's friend and potentially end up at Merleau-Ponty's hands.  There may be overlap with the previous expressions, or there may not, but the exercise should at least give more potential paths for us to consider.

\section{From Wigner's Friend \ldots}    \label{Wigner's Hands}

Wigner first published his famous thought experiment in a 1961 volume titled {\sl The
Scientist Speculates}~\cite{Wigner1961}.  Indeed, what a speculation it was!
Sixty-two years later, it is still a hot topic of discussion~\cite{Frauchiger2018,Baumann2020,Cavalcanti2021}, with even a well-funded workshop devoted to it in 2022~\cite{Zeng2022}.  Most interesting from our perspective, however, is how the dissolution of the conundrum's paradoxical character involves all eight QBist tenets.  At the same time it showcases a further feature of QBist thinking that has yet to be discussed.

To contrast with QBism's story, let us refer to Wigner's original description of his conundrum: We will quote it at length so that no details are missed.\footnote{Though, in place of his original mathematical expressions we will use Dirac notation and some notational shortcuts for ease of reading.} Following his convention, we will speak of ``the observer'' and ``the friend'' in all our commentary.\footnote{As well, we will disambiguate those instances in his text where he writes ``observer'' when he clearly means ``friend.''} Finally, we highlight in boldface those places where QBism has some disagreement with Wigner's account of the thought experiment.

First off, Wigner makes it clear that he intends to think in epistemic terms about quantum states.  (It is another matter whether he does so consistently.)  Indeed, Wigner starts off in a distinctly QBist direction.  He writes,
\bq
Given any object, all possible {\bf knowledge concerning that object}$^{(a)}$
can be given as its wave function. \ldots\ If one knows [the wave function], one can
foresee the behavior of the object as far as it {\it can\/} be foreseen. More precisely,
the wave function permits one to foretell with what probabilities
the object will make one or another impression on us if we let it interact
with us either directly, or indirectly. \ldots

One realises that {\it all\/} the information which the
laws of physics provide consists of probability connections between subsequent
impressions that a system makes on one {\bf if one interacts with it repeatedly},$^{(b)}$ \ldots. The wave
function is a convenient summary of that part of the past impressions
which remains relevant for the probabilities of receiving the different
possible impressions when interacting with the system at later times.
\eq
Note that his probabilities are not about the true states of the world, but about ``impressions'' or ``sensations.''  This is on a trajectory toward QBism's ``experiences,''\footnote{See for instance \cite{Moller2008} for some discussion of James's ``rich notion of experience.'' Until our recent readings of Merleau-Ponty, this has been the historical model QBism has followed.} but the divergence between Wigner's understanding of quantum theory and QBism start to become apparent in the middle ellipsis above:
\bq
\noindent The information given by the wave function is communicable.  {\bf If
someone else somehow determines the wave function of a system, he
can tell me about it and, according to the theory, the probabilities for
the possible different impressions (or ``sensations'') will be equally large,
no matter whether he or I interact with the system in a given fashion.
In this sense, the wave function ``exists.''}$^{(c)}$
\eq
Next, as a preamble to his paradox, Wigner starts with a general question about what if an observation is made ``indirectly'':
\bq
It is natural to inquire about the situation if one does not make the
observation oneself but lets someone else carry it out. {\bf What is the wave
function if my friend}$^{(d)}$ looked at the place where the flash might show
at time $t$? The answer is that the information available about the {\it object\/}
cannot be described by a wave function. {\bf One could attribute a wave
function}$^{(e)}$ to the joint system: friend plus object, and this joint system
would have a wave function also after the interaction, that is, after my
friend has looked. I can then enter into interaction with this joint system
by asking my friend whether he saw a flash. If his answer gives me
the impression that he did, the joint wave function of friend
$+$ object will change into one in which they even have separate wave functions
(the total wave function is a product) and the wave function of the
object is $|\psi_1\rangle$. If he says no, the wave function of the object is
$|\psi_2\rangle$ i.e., the
object behaves from then on as if I had observed it and had seen no
flash. However, even in this case, in which the observation was carried
out by someone else, the typical change in the wave function occurred
only when some information (the {\it yes\/} or {\it no\/} of my friend) entered
{\it my\/} consciousness. It follows that the quantum description of objects is
influenced by impressions entering my consciousness.
\eq
With that much as an introduction, Wigner gets to what he sees as the heart of the matter:
\bq
Does \ldots\ consciousness influence the physico-chemical
conditions? In other words, does the human body deviate from
the laws of physics, as gleaned from the study of inanimate nature? \ldots\
[A]t least two reasons can be given to support [a ``Yes'' to this]. \ldots

In order to present [the first] argument, it is necessary to follow my description
of the observation of a ``friend'' in somewhat more detail than was
done in the example discussed before. Let us assume again that the object
has only two states, $|\psi_1\rangle$ and $|\psi_2\rangle$.  If the
state is, originally, $|\psi_1\rangle$
the state of object plus [friend] will be, after the interaction,
$|\psi_1\rangle|\chi_1\rangle$; if the state of the object is $|\psi_2\rangle$,
the state of object plus [friend] will be $|\psi_2\rangle|\chi_2\rangle$
after the interaction. The wave functions $|\chi_1\rangle$ and $|\chi_2\rangle$
give the state of the [friend]; in the first case {\bf he is in a state}$^{(f)}$ which responds to the question
``Have you seen a flash?''\ with ``Yes''; in the second state, with ``No.''
There is nothing absurd in this so far.

Let us consider now an initial state of the object which is a linear
combination $\alpha|\psi_1\rangle+\beta|\psi_2\rangle$ \ldots\@.
It then {\it follows\/} from the linear nature of the quantum mechanical equations of motion that
the state of object plus [friend] is, after the interaction,
\normalsize
\be
|\Phi\rangle = \alpha |\psi_1\rangle|\chi_1\rangle + \beta
|\psi_2\rangle|\chi_2\rangle\;.
\label{CheeCheePah}
\ee
{\bf If I now ask the [friend] whether he saw a flash, he will with
a probability $|\alpha|^2$ say that he did, and in this case the object will also
give to me the responses as if}$^{\,(g)}$ it were in the state $|\psi_1\rangle$.
If the [friend] answers ``No''---the probability for this is $|\beta|^2$---the object's responses from
then on will correspond to a wave function $|\psi_2\rangle$.
{\bf The probability is zero}$^{\,(h)}$ that the [friend] will say ``Yes,'' but the object gives the response which
$|\psi_2\rangle$ would give because the wave function $|\Phi\rangle$
of the joint system has no $|\psi_2\rangle|\chi_1\rangle$ component. Similarly, if the [friend] denies
having seen a flash, the behavior of the object cannot correspond to
$|\chi_1\rangle$ because the joint wave function has no $|\psi_1\rangle|\chi_2\rangle$
component. All this is quite satisfactory: {\bf the theory of measurement, direct or indirect, is
logically consistent so long as I maintain my privileged position as ultimate
observer.}$^{(i)}$

However, if after having completed the whole experiment I ask my
friend, ``What did you feel about the flash before I asked you?''\ he will
answer, ``I told you already, I did (did not) see a flash,'' as the case may
be. In other words, the question whether he did or did not see the
flash was already decided in his mind, before I asked him.
If we accept this, {\bf we are driven to the conclusion that the proper wave
function}$^{(j)}$ immediately after the interaction of friend and object was already
either $|\psi_1\rangle|\chi_1\rangle$ or $|\psi_2\rangle|\chi_2\rangle$ and not the linear
combination $|\Phi\rangle$.  This
is a contradiction, because the state described by the wave function
$|\Phi\rangle$ describes a state that has
properties which neither
$|\psi_1\rangle|\chi_1\rangle$ nor $|\psi_2\rangle|\chi_2\rangle$
has. If we substitute for ``friend'' some simple physical apparatus, such as an atom
which may or may not be excited by the light-flash, this difference has observable effects and {\it there is no doubt
that $|\Phi\rangle$ describes the properties of the joint system correctly, the assumption that the wave
function is either $|\psi_1\rangle|\chi_1\rangle$ or $|\psi_2\rangle|\chi_2\rangle$ does
not}. {\bf If the atom is replaced by a conscious being, the
wave function $|\Phi\rangle$ \ldots\ appears absurd because it
implies that my friend was in a state of suspended animation before he answered
my question.}$^{(k)}$

It follows that the being with a consciousness must have a different role in quantum mechanics than
the inanimate measuring device: the atom considered above.
In particular, {\bf the quantum mechanical equations of motion cannot be
linear if the preceding argument is accepted.}$^{(l)}$  This argument implies that ``my
friend'' has the same types of impressions and sensations as I---in particular,
that, after interacting with the object, he is not in that state of suspended
animation which corresponds to the wave function $|\Phi\rangle$.

It is not necessary to see a contradiction here from the point of view of orthodox quantum mechanics,
and there is none if we believe that the alternative is meaningless,
whether my friend's consciousness contains either the impression of having seen a flash or of not having seen a flash. However, {\bf to deny
the existence of the consciousness of a friend to this extent is surely an
unnatural attitude, approaching solipsism, and few people, in their
hearts, will go along with it.}$^{(m)}$
\eq

Let us now comment on the highlighted parts.
\begin{enumerate}[(a)]
\item
{\bf ``Knowledge.''} Note that nowhere in Wigner's description does he deny the epistemic character of quantum states.  In fact, that they are epistemic is the bedrock of his considerations, even if he is not exactly consistent about it. However, he differs from QBism in that his epistemic states are about knowledge, not belief. This gives them a factive character that QBism denies. For if it were so, a quantum state could be right or wrong as set by the facts of the world.  In the development of Wigner's argument, this distinction will play a pivotal role.

\item
{\bf ``Probability connections \ldots\ if one interacts with it repeatedly?''} He may be thinking of probability in a frequentist rather than a Bayesian conception here, but it is not clear.  If he is, it would help shore up why he is taking a factive notion of quantum states.

\item
{\bf ``If someone determines the wave function of a system, he can tell me about it and the probabilities for the possible sensations [for each of us] will be equally large, no matter whether he or I interact with the system.'' In this sense, the wave function exists.}  As Wigner says, {\it if so}, this would make a quantum state exist in a way contrary to de Finetti's dictum that ``probability does not exist.'' QBism instead requires a much more nuanced discussion of this situation, say along the lines of that described in Section~\ref{SingleUser}.  If you were to tell me your quantum state for a system, I might or might not take it on myself.  Or, I might or might not roll it in to my previous quantum-state assignment and end up with something completely new. What is for sure is that in QBism a quantum state assignment is not blindly transferable.

\item
{\bf ``What is the wave function if my friend \ldots.''}  Notice the definite article here, {\it the}.  The very tone of the question eschews the QBist conception of quantum states as pluralistic and always tagged to a specific agent.  For QBism, there might be as many quantum states for a single quantum system as there are agents considering it.

\item
{\bf ``One could attribute a wave function \ldots.''}  This is the same sin as reported in the last item.  Who is the ``one''?  QBism never tolerates ambiguity on this point, but Wigner to some extent flicks back and forth between the two players in the story.  It appears the narrative is moving toward making the observer a super-observer who in the end must be made consistent with any inside observers (eg., the friend).  But remember, Wigner had already settled on states being epistemic: Why then demand identity between the two observers' state assignments?  Presumably because then one observer would otherwise get the facts wrong.

\item
{\bf ``The wave functions \ldots\ give the state of the observer; in the first case he is in a state \ldots.''}  Notice how perilously close this gets to an ontic-tinged notion of quantum states.  He is {\it in\/} a state?  It makes states sound like properties that objects can have. This is a slippery slope to an irreversible confusion between epistemic and ontic.

\item
{\bf ``If I ask the observer whether he saw a flash, he will with some probability say that he did, and in this case the object will also
give to me the responses as if it were in the state \ldots.''}  As if?  This tentatively privileges the state of the observer in the narrative.  This will drive the idea that the observer's state assignment must ultimately be made consistent with the friend's, so to erase the ``as if.''

\item
{\bf ``The probability is zero \ldots.''}  It is hard to know exactly how Wigner is thinking of probability-zero, but I would guess he identifies it with a proposition's truth value being ``false.''

\item
{\bf ``The theory of measurement is logically consistent so long as I maintain my privileged position as ultimate observer.''}  Of course, QBism argues that the theory is consistent precisely because it {\it does not privilege\/} any agents's personal point of view.  But then, it does that in a very different way than Wigner is aiming for.

\item
{\bf ``We are driven to the conclusion that the proper wave function \ldots.''}  When quantum states are factive to begin with, then sentences like this might have a meaning.  But QBism undermines that; it says that it was always a mistake to think there is a single ``proper wave function'' across distinct agents.

\item
{\bf ``If the atom is replaced by a conscious being, the entangled wave function would imply my friend was in a state of suspended animation before he answered my question.''}  What might he mean by ``suspended animation'' here?  Presumably he means that the friend has no experiences while the observer holds the state $|\Phi\rangle$.  But why on earth would it imply that?  $|\Phi\rangle$ was supposed to only ``foretell with what probabilities the object will make one or another {\it impression\/} on us'' [our emphasis]. The extra supposition of ``suspended animation'' for the friend violates Wigner's very starting point.

\item
{\bf ``The quantum mechanical equations of motion cannot be linear if the preceding argument is accepted.''}  It is interesting to think about what seems to be going on here.  Wigner accepts outright that quantum states are epistemic, but then he demands a dualistic intervention of matter on mind without mind gaining any information in the process. If not completely so or consistently so with quantum states, Wigner is surely thinking in ontic terms for time evolution processes.

\item
{\bf ``To deny the existence of the consciousness of a friend to this extent is surely an unnatural attitude, approaching solipsism, and few people, in their hearts, will go along with it.''}  Again, why would the observer's assignment $|\Phi\rangle$ to the composite system amount to ``denying the existence of the consciousness of the friend?'' Diagnosis: Likely because, though Wigner claims $|\Phi\rangle$ to be epistemic about the observer's potential sensations, he is not so consistent about it. He also seems to hold that $|\Phi\rangle$ is directly representative of what is going on in the closed system as well.  Treatment: QBism, of course.  So, let's get to it.
\end{enumerate}

\begin{figure}
    \begin{center}
    \includegraphics[width=\linewidth]{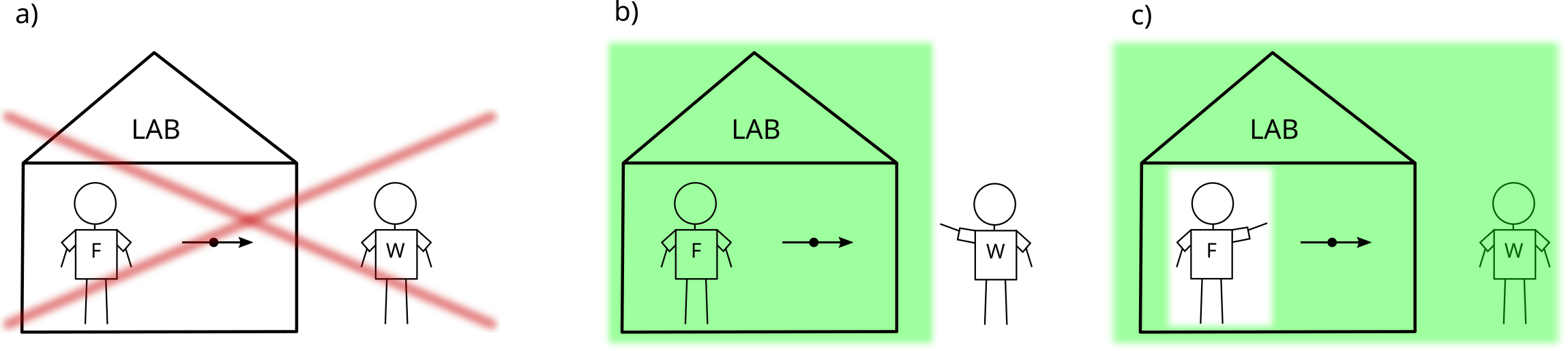}
\end{center}
\caption{\label{ThreeHouses} (a) In usual descriptions of various Wigner's-friend thought experiments, there is an urge to portray everything from a God's eye view. Here, we depict the thought experiment, where Wigner, his friend, and a spin-1/2 particle interact, and we symbolize QBism's disapproval of such portrayals with a big red X\@. In QBism, the quantum formalism is only used by agents who stand {\it within\/} the world; there is no God's-eye view. (b) Instead, in QBism, to make predictions, Wigner treats his friend, the particle, and the laboratory surrounding her (all shaded in green) as a physical system external to himself. While (c) to make her own predictions, the friend must reciprocally treat Wigner, the particle, and her surrounding laboratory (all shaded in green) as a physical system external to herself. It matters not that the laboratory spatially surrounds the friend; it, like the rest of the universe, is external to her agency, and that is what counts.}
\label{}
\end{figure}

In contrast to the way Wigner sets up the problem, QBism sees the thought experiment as being about two equal-footed agents, each of which may take actions on the system of their concern and receive deeply personal experiences as a consequence.  See Fig.\ \ref{ThreeHouses}. We know that Wigner is not treating the agents on an equal footing because everything in his story is expressed univocally from the perspective of the outside observer.  It is the outside observer who specifies what quantum state the friend will hold for the object after seeing a flash or not.  But does anyone ever ask the friend how he would gamble on the sensations he would receive by taking actions on the outside observer instead?  Indeed, we know the outside observer is a user of quantum mechanics\footnote{Remember the definitions from Section~\ref{KikiMowing}.} rather than an uneducated agent, as he writes down quantum states, unitary operations, and makes probability assignments through the Born rule. Yet Wigner never bothers to tell us whether the friend too is a user of quantum mechanics.  Wigner could have played the ``as if'' game mentioned in comment (g) all the way down, with $|\psi_1\rangle$ and $|\psi_2\rangle$ never being the friend's actual state assignments, just the ones the observer thinks the friend {\it ought to\/} make.  The states $|\chi_1\rangle$ and $|\chi_2\rangle$ stand only for the answers to the dumb query, ``Have you seen the flash?'' The friend could answer that without knowing a bit about quantum mechanics.

Thus it is crucial for QBism's conception of quantum theory as a user's manual that it treats the two agents on an equal footing.  Furthermore, QBism cannot restrict the applicability of quantum mechanics to inanimate objects, else the manual would not be universal to everything external to the agent. As already broached in Section \ref{EveningBreeze}, an agent can even apply the normative quantum calculus to their expectations arising from actions they might take on other agents.

Yet QBism claims that despite Wigner, all of this can be done without modifying quantum mechanics in any way.  As already emphasized, it takes all the apparatus of QBism to fight off the various quantum conundrums.  But one might ask really what are the most salient points for this {\it particular\/} conundrum?  There are three.

First is the idea that quantum-state assignments do not ``pierce into'' the systems to which they are ascribed.  They do not describe what is ``going on in there.'' Their sole role is to capture the ascriber's gambling attitudes toward the consequent personal experiences that would arise from actions on the system---this is true of any quantum state $|\psi\rangle$. Consequently it is no less true when it comes to state assignments like $|\Phi\rangle$ in Eq.\ (\ref{CheeCheePah}). The latter may look more like it is talking about the goings on inside, but if so, that is surely an artefact of the representation.  Every term in it is about the outside observer's beliefs concerning his own experiences.  There is no metaphysical necessity that the friend even make the state assignments $|\psi_1\rangle$ and $|\psi_2\rangle$ himself, and QBism in fact basks in the possibility that it might be so.  In a slogan, ``no matter what quantum state assignment is made to a system, the system is more than that.''  The outside observer might make the assignment $|\Phi\rangle$ and yet the friend decide to make no measurement at all \ldots\ or even do something completely absurd like {\it eat the object\/} and incorporate it into himself!  If a simple spin can surprise a measuring agent by shunning its probability-1 for the outcome of some observable, how much more so might another full-blown agent?  But this was already emphasized in Section~\ref{EveningBreeze}.

The second has to do with the idea of ``suspended animation.'' There is a tendency in many if not most discussions concerned with assigning quantum states to agents that pure-state assignments somehow contradict the very idea of agency.  After all, if the assigner has made a pure-state assignment he must be in a position to apply any unitary operation he wishes to make the state anything he will.  For instance, in the implied quantum evolution of the friend plus object, one has
\be
|\Phi_0\rangle = \Big(\alpha |\psi_1\rangle + \beta
|\psi_2\rangle\Big)|\chi_0\rangle
\qquad\overset{\hat U}{\longrightarrow}\qquad
|\Phi\rangle = \alpha |\psi_1\rangle|\chi_1\rangle + \beta
|\psi_2\rangle|\chi_2\rangle\;,
\label{Chalupah}
\ee
where $|\chi_0\rangle$ is a ``ready'' state for the friend.  One says that the left state $|\Phi_0\rangle$ represents that the friend's measurement has yet to happen, whereas the right state $|\Phi\rangle$ represents that the process is complete.  But then the unitary operation $\hat U$ that took the process forward can be reversed by $\hat U^{-1}$ to take it all back---so that the measurement outcome goes ``poof,'' right back out of existence.  For instance as already quoted in Section~\ref{SuperBowlSunday}, Baumann and Brukner \cite{Baumann2020} write that $\hat U$ ``is determined by the interaction Hamiltonian between the friend and the system, which models the measurement and is assumed to be known to and in [the] control of Wigner.''  However, one must remember that like a quantum state, within QBism $\hat U$ is purely a compendium of beliefs, and itself does not pierce into the metaphysical goings on the friend's closed laboratory.  $\hat U^{-1}\hat U=I$ is purely a statement about the outside observer's expected experiences, not about what is happening out in the world.

In fact the disparity between QBism and the usual discussions goes deeper:  If a QBist agent ever works up a state of belief that corresponds to assigning a pure quantum state to her friend, by necessity, it will be precisely because she believes him to be another free-willed agent such as herself, not in contradiction to it.  Take some {\it system\/} that our QBist agent believes will pass a Turing test should she administer it, will pass Danny Greenberger's ``large red button test''~\cite{Greenberger2014}, one for which she feels a heartfelt empathy, one for which she would feel the loss of their companionship should they leave, etc., etc.  One can consider a very long list of such things.  If they are all things she believes, then they are all things she believes: They must be rolled into her quantum state assignment for the system. They are not thoughts to be ignored, but rather ones to be refined until they fit within a numerical framework capturing the agent's best gambles.  So, what if that rolling up (along with a myriad of other details of the agent's belief system) leads to some pure-state assignment?  QBism's retort is, ``Indeed, so what? There's nothing to be made of it.''  It only means that the pure-state assignment landed on will contain {\it all\/} those beliefs.  Rather than contradict anything, all those beliefs are necessarily in there.

So, $|\Phi\rangle$ $=$ a state of suspended animation? {\it Pah!}  But then that leaves us with the toughest issue of all: This is the third salient point alluded to above, and it is the most profound, for it takes us into the territory of ontology.  For every agent in the Wigner's friend story there is something that happens---new experience comes into existence.  Yet, the quantum formalism itself is always thoroughly and exclusively first personal in any particular application.  Whatever is gambled on is always ``my experience,'' whoever the ``me'' happens to be. This is the ``QBist Copernican principle'' first proto-formulated in~\cite{Fuchs10a,Fuchs13a,Fuchs12,Fuchs2014} and perhaps most thoroughly elaborated in~\cite{Fuchs2017}.  See also~\cite{Cavalcanti2021} for a sympathetic analysis from an independent point of view.

Yet, how can this be?  Wigner's quantum state assignment cannot pierce into his friend's reality, but he can still maintain this Copernican principle.  If this feels like a contradiction, it shouldn't. The state assignment lives exclusively at the level of the first personal, while the general structure of quantum theory with its normative suggestions lives at the level of the third-personal.  We called it normative structural realism before.  If quantum theory is a user's manual that any agent can use, then the presupposition for all agents is that they shall experience.  From this point of view, the lesson of quantum theory is that {\it experience happens}, and the refinement of the notion that comes from Wigner's thought experiment is that there {\it is a sense\/} in which those experiences need not live in a single universe.

Here is the way we put it in~\cite{Fuchs12}:
\bq
\noindent
The only glaringly mutual world there is for Wigner and his
friend in a QBist analysis is the partial one that might come about if these two
bodies were to later take actions upon each other (``interact'')---the rest of the
story is deep inside each agent’s private mesh of experiences, with those having
no necessary connection to anything else.

But what a limited story [Wigner's friend] is: For its concern is only of agents and the
systems they take actions upon. What we learn from Wigner and his friend is
that we all have truly private worlds in addition to our public worlds. But QBists
are not reductionists, and there are many sources of learning to take into account
for a total worldview---one such comes from Nicolas Copernicus: That man should
not be the center of all things (only some things). Thus QBism is compelled as
well: What we have learned of agents and systems ought to be projected onto all
that is external to them too. The key lesson is that each part of the universe has
plenty that the rest of the universe can say nothing about. That which surrounds
each of us is more truly a pluriverse.
\eq
This always meshed well with William James's ``republican banquet'' vision of a pluriverse~\cite{James1882}:
\bq
\noindent
Why may not the world be a sort of republican banquet of this sort, where all the qualities of being respect one another's personal sacredness, yet sit at the common table of space and time?

To me this view seems deeply probable.  Things cohere, but the act of cohesion itself implies but few conditions, and leaves the rest of their qualifications indeterminate. \ldots\

[I]f we stipulate only a partial community of partially independent powers, we see perfectly why no one part controls the whole view, but each detail must come and be actually given, before, in any special sense, it can be said to be determined at all.  This is the moral view, the view that gives to other powers the same freedom it would have itself.
\eq

But the phenomenology of Merleau-Ponty gives us heart that the last word has still not been said on this.  For all that we have learned from the Wigner's friend thought experiment, we now start to see that the standard diagramming of the story, with its spin-1/2 particle, a laboratory surrounding the friend, etc., is just so much distraction.  The same holds for QBism's solution as depicted in Fig.\ \ref{ThreeHouses}.  Perhaps a more essential point can be found by stripping away these extra elements and focusing on Wigner and the friend {\it tout court}. What further insight might be gleaned from a simple handshake between Wigner and his friend?

\section{\ldots\ to Merleau-Ponty's Hands, Concluding Remarks}
\label{DuhDuhDuh}

John Bell once asked, ``[A]re we not obliged to admit that more or less `measurement-like' processes are going on more or less all the time more or less everywhere?''~\cite{Bell1990}  It will come as a surprise to many in the philosophy of physics community that in fact QBism agrees heartily with Bell on this point.  The words will have a different intention than Bell's, but indeed, {\it yes we are obliged!} However, as such we genuinely do need to say more about what ``measurement-like'' is.  This much has been settled:  When we are talking about an application of the quantum formalism concerning an agent taking an action on a system, the ``measurement outcome'' just is the consequent experience.  So ``measurement-like process'' $=$ ``lived experience'' in this case.  The suggestion then is that this applies more broadly, that experience\footnote{I say ``experience'' here, but more work should be done to find the right word for this de-anthropocentrized ontological distillate of QBist considerations.  Other potential terms are pure experience~\cite{James96a}, neutral monism~\cite{Banks2014,Atmanspacher2022}, phenomenon~\cite[pp.\ 178--192]{Plotnitsky2021}, elementary quantum phenomenon~\cite{Wheeler83b}, actual occasion~\cite{Whitehead1978}, intra-action~\cite{Barad2007}, trans-action~\cite{Dewey1989}, chiasm~\cite{Merleau-Ponty1968}, flesh~\cite{Merleau-Ponty1968}, or even QBoom~\cite[pp.\ 2193--2194, 2213--2214]{Samizdat2}.}  is somehow the very stuff of the world~\cite{Fuchs2017}.

At first sight, it might appear that this move is of a spectrum with the various ``perspectival'' or ``relational'' ontologies being explored for quantum theory currently~\cite{Barzegar2022,Brukner2022}---for instance in Rovelli's relational quantum mechanics (RQM)~\cite{Rovelli1996}, Brukner's relative facts~\cite{Brukner2018}, Healey's quantum pragmatism~\cite{Healey2017}, or Glick's perspectival normative realism~\cite{Glick2021}.  But if one looks closely at the apparatus expounded in this paper, one will see that it's just not the case.  The very words ``perspectival'' or ``relational'' evoke a block-universe conception of the world, a world which is in some sense already complete and unified. That for instance might be part of the psychology for the recent retreat of Adlam and Rovelli~\cite{Adlam2022} from a more radical version of RQM~\cite{Pienaar2021a,Pienaar2021b}.
Compare that kind of thinking to this passage from QBist lore~\cite{Fuchs2007}:
\bq
[I]f anything, the Bayesian account of quantum theory [now known as QBism] is essentially the opposite of solipsism. Rather than a unity to nature, it suggests a plurality. An image that might be useful (but certainly flawed) comes from Escher's various paintings of impossible objects. The viewer would initially like to think of them as 2D projections of a three-dimensional object; but he cannot. Now imagine how much worse it would get if we were to have two viewers with two slightly different paintings, each purporting to be a different perspective on ``the'' impossible object. Since neither viewer can lift from his own 2D object to a 3D one, there is no way to unify the pictures into a single whole.
\eq

So perspectivalism really does not work when it comes to QBism's ontological project.  Experience is a far richer notion than a perspective on on some pre-existing thing. As well it is far more than a simple relation among pre-existing things.  Lived experience has an autonomy that neither of these notions capture.  Each quantum measurement creates something new in the universe that is above and beyond the agent's relation to the quantum system they are acting upon.  Quantum measurement in QBism is more like childbirth.  Without a father and mother, there would be no child, but the child does not express merely the mother's relation to the father or vice versa.  The child is something new and {\it sui generis}.  To be ``measurement-like''---in QBism's retort to John Bell---is to be like a fresh moment of creation not unakin to what is imagined of the big bang itself.\footnote{See the discussion between John Wheeler and theologian Richard Elvee in \cite{Wheeler1982}:

ELVEE: Dr.\ Wheeler, who was there to observe the universe when it
started? Were we there? Or does it only start with our observation?
Is the big bang here?

WHEELER: A lovely way to put it---``Is the big bang here?'' I can
imagine that we will someday have to answer your question with a
``yes.'' \ldots\ Each elementary quantum phenomenon is an elementary act of ``fact
creation.'' That is incontestable. But is that the only mechanism
needed to create all that is? Is what took place at the big bang the
consequence of billions upon billions of these elementary processes,
these elementary ``acts of observer-participancy,'' these quantum
phenomena? Have we had the mechanism of creation before our eyes all
this time without recognizing the truth? That is the larger question
implicit in your comment.}

\begin{figure}
\begin{center}
\includegraphics[width=5.7in]{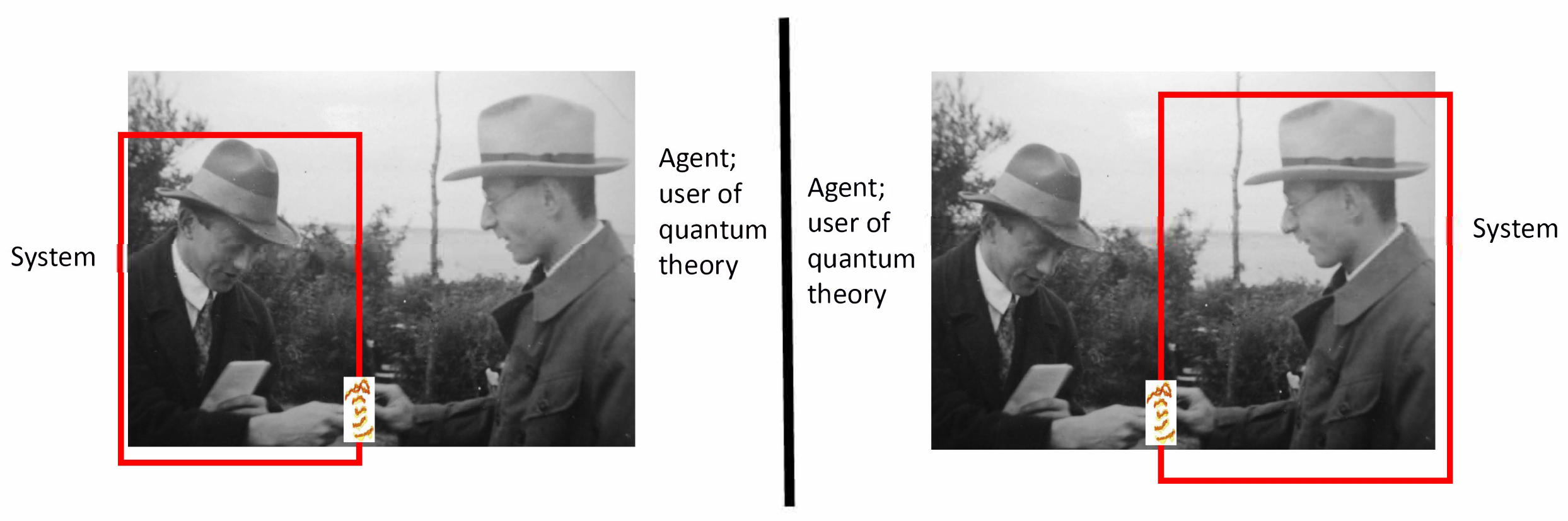}
\end{center}
    \caption{\label{WignerAndFriend} Merleau-Ponty's hands, upgraded to the extreme.  Eugene Wigner (right) and his friend Werner Heisenberg (left) passing a pen between themselves.  One might as well say they are shaking hands. There is a ``tingly bit'' that, even if it cannot be expressed mathematically across the two agents' uses of the theory, must somehow be treated symmetrically.  After all, Wigner and friend both are both agent and both system; it is only the story as related from one focus or the other that appears asymmetrical.}
\end{figure}

Maurice Merleau-Ponty used reflections on a person touching one hand with the other to develop his ontology of chiasm and flesh.  Here is a sample of the thinking~\cite[pp.\ 140--142]{Merleau-Ponty1968}:\footnote{I apologize for introducing this lengthy quote, but it seemed important to have it displayed here and now. Nonetheless, it comes from a man who was in deep need of some paragraphing lessons.}
\bq
\noindent
If we can
show that the flesh is an ultimate notion, that it is not the union
or compound of two substances, but thinkable by itself, if there
is a relation of the visible with itself that traverses me and
constitutes me as a seer, this circle which I do not form, which
forms me, this coiling over of the visible upon the visible, can
traverse, animate other bodies as well as my own. And if I was
able to understand how this wave arises within me, how the
visible which is yonder is simultaneously my landscape, I can
understand a fortiori that elsewhere it also closes over upon
itself and that there are other landscapes besides my own. If it
lets itself be captivated by one of its fragments, the principle of
captation is established, the field open for other Narcissus, for
an ``intercorporeity.'' If my left hand can touch my right hand
while it palpates the tangibles, can touch it touching, can turn its
palpation back upon it, why, when touching the hand of another,
would I not touch in it the same power to espouse the things that
I have touched in my own? It is true that ``the things'' in question
are my own, that the whole operation takes place (as we say) ``in
me,'' within my landscape, whereas the problem is to institute
another landscape. When one of my hands touches the other, the
world of each opens upon that of the other because the operation
is reversible at will, because they both belong (as we say) to one
sole space of consciousness, because one sole man touches one
sole thing through both hands. But for my two hands to open
upon one sole world, it does not suffice that they be given to one
sole consciousness---or if that were the case the difficulty before
us would disappear: since other bodies would be known by me in
the same way as would be my own, they and I would still be
dealing with the same world. No, my two hands touch the same
things because they are the hands of one same body. And yet
each of them has its own tactile experience. If nonetheless they
have to do with one sole tangible, it is because there exists a very
peculiar relation from one to the other, across the corporeal
space---like that holding between my two eyes---making of my
hands one sole organ of experience, as it makes of my two eyes
the channels of one sole Cyclopean vision. \ldots\ Now why would this generality, which constitutes the unity of my body, not open it to other bodies? The handshake too is reversible; I can feel myself touched as well and at the same time as touching \ldots\ Why would not the synergy exist among different organisms, if it is possible within each? Their landscapes interweave, their actions and their passions fit together exactly: this is possible as soon as we no longer make
belongingness to one same ``consciousness'' the primordial definition of sensibility, and as soon as we rather understand it as the return of the visible upon itself, a carnal adherence of the sentient to the sensed and of the sensed to the sentient. For, as overlapping and fission, identity and difference, it brings to birth a ray of natural light that illuminates all flesh and not only my own.
\eq
It is pretty clear that this argument begs to be compared to the touch shared between Wigner and his friend in Fig.\ \ref{WignerAndFriend}.  Could we pick up where Merleau-Ponty left off?  For though M-P ends up with a metaphysics of ``radical contingency''~\cite{Bertram1980}, just as QBism itself does~\cite{Fuchs2017,Samizdat2,Fuchs2007}, QBism still has something else to say.  {\sl Taken in turn}, each side of the divide in Fig.\ \ref{WignerAndFriend} tolerates the addition of the normative quantum formalism.  Why should that be?  What does it tell us further about nature that the reversibility of the touch does not?

There can be no answer to these questions here, but at least now we know where to look.  Moreover, there is already some good material in the literature to give us a start, some in this volume and some elsewhere~\cite{Schack2022,Bitbol2020,Bitbol2021,Bitbol2023,vonBaeyer2023}.  As emphasized in the Introduction, Berghofer and Wiltsche ask, ``The question is not whether QBists and phenomenologists should attempt to join forces, but what has taken us so long?''  Indeed, let us forge this alliance {\it now}, for the battle will be hard! \medskip

\begin{flushright}
\baselineskip=13pt
\parbox{3.1in}{\baselineskip=13pt
     This story shall the good man teach his son;
\\      And Crispin Crispian shall ne'er go by,
\\      From this day to the ending of the world,
\\      But we in it shall be remembered---
\\      We few, we happy few, we band of brothers;
\\      For he to-day that sheds his blood with me
\\      Shall be my brother; be he ne'er so vile,
\\      This day shall gentle his condition;
\\      And gentlemen in England now-a-bed
\\      Shall think themselves accurs'd they were not here,
\\      And hold their manhoods cheap whiles any speaks
\\      That fought with us upon Saint Crispin's day.
}\bigskip\\
--- Henry V
\end{flushright}


\section*{Acknowledgements}

First and foremost I thank those colleagues I called the ``few exceptions'' in the Introduction. Despite my damning statement, I have had many pleasant conversations with Guido Bacciagaluppi, Florian Boge, Harvey Brown, Jeff Bub, David Glick, Richard Healey, Jenann Ismael, Huw Price, Chris Timpson, Jos Uffink, and David Wallace. I also cherish what I learned from my friends Bill Demopoulos, Itamar Pitowsky, and Pat Suppes who are no longer with us.  None of these colleagues managed to back me away from what they perceived as the dangers of QBism, but they respected the subject enough to have a conversation. Next, I thank Mario Hubert for a discussion that helped sow the seed of normative structural realism as descriptive of one aspect of QBism. Huge gratitude goes to Amanda Gefter for the generous sharing of her unpublished thoughts, which make their first public appearance here. Finally, I graciously thank the Institute for Quantum Optics and Quantum Information (IQOQI) in Vienna for its hospitality during a stage of the writing of this article. My academic great-great-great-grandfather Franz Exner, with his belief in indeterminism even before quantum mechanics, walked the same streets, and I feel his spirit looks in on QBism whenever I am there. This work was supported in part by National Science Foundation Grant 2210495 and in part by Grant 62424 from the John Templeton Foundation. The opinions expressed in this publication are those of the author and do not necessarily reflect the views of the John Templeton Foundation.


\begin{thebibliography}{999}

\bibitem{Fuchs10a}
C.~A. Fuchs, ``QBism, the Perimeter of Quantum Bayesianism,'' \href{https://arxiv.org/abs/1003.5209}{\tt arXiv:1003.5209}.

\bibitem{Fuchs13a}
C.~A. Fuchs and R.~Schack, ``Quantum-Bayesian Coherence,'' Rev.\ Mod.\ Phys.\ {\bf 85}, 1693--1715 (2013); \href{https://arxiv.org/abs/1301.3274}{\tt arXiv:1301.3274}.

\bibitem{Fuchs14a}
C.~A. Fuchs, N. D. Mermin, and R.~Schack, ``An Introduction to QBism with an Application to the Locality of Quantum Mechanics,'' Am.\ J. Phys.\ {\bf 82}, 749--754 (2014); \href{https://arxiv.org/abs/1311.5253}{\tt arXiv:1311.5253}.

\bibitem{Fuchs2017}
C. A. Fuchs, ``Notwithstanding Bohr, the Reasons for QBism,'' Mind and Matter {\bf 15}, 245--300 (2017);  \href{https://arxiv.org/abs/1705.03483}{\tt arXiv:1705.03483}.

\bibitem{Caves02a}
C.~M. Caves, C.~A. Fuchs and R.~Schack, ``Quantum Probabilities as Bayesian Probabilities,'' Phys.\ Rev.\ A {\bf 65}, 022305 (2002); \href{https://arxiv.org/abs/quant-ph/0106133}{\tt arXiv:quant-ph/0106133}.

\bibitem{Mischel2014}
W. Mischel, {\sl The Marshmallow Test:\ Why Self-Control Is the Engine of Success}, (Little, Brown and Company, New York, 2014).

\bibitem{Fuchs2002b}
C. A. Fuchs, \href{http://www.physics.umb.edu/Research/QBism/WHAT.pdf}{\sl Quantum States:\ What the Hell Are They?}, unpublished (2002).

\bibitem{Gefter2014}
A. Gefter, {\sl Trespassing on Einstein's Lawn:\ A Father, a Daughter, the Meaning of Nothing, and the Beginning of Everything}, (Bantam Books, New York, 2014).

\bibitem{Gefter2015}
A. Gefter, ``\href{https://www. quantamagazine.org/20150604-quantum-bayesianism-qbism/}{A Private View of Quantum Reality},'' {\sl Quanta Magazine}, 4 June 2015.

\bibitem{Gefter2022}
A. Gefter, ``Amanda Gefter on Schr\"odinger's QBist cat,'' in ``\href{https://tinyurl.com/mr4cfr32}{The many meanings of Schrödinger's cat},'' IAI News, 24 February 2022.

\bibitem{Durant2006}
W. Durant, {\sl The Story of Philosophy:\ The Lives and Opinions of the Greater Philosophers}, (Pocket Books, New York, 2006).

\bibitem{Bohr1962}
Interview of Niels Bohr by T.~S. Kuhn, L.~Rosenfeld, A.~Petersen, and E.~Rudinger on 1962 November 17, Niels Bohr Library \& Archives, American Institute of Physics, College Park, MD USA,
\href{https://www.aip.org/history-programs/niels-bohr-library/oral-histories/4517-5}{\tt www.aip.org/history-programs/niels-bohr-library/oral- histories/4517-5}.

\bibitem{Caves02b}
C.~M. Caves, C.~A. Fuchs and R.~Schack, ``Unknown Quantum States:\ The Quantum de Finetti Representation,'' J. Math.\ Phys.\ {\bf 43}, 4537--4559 (2002); \href{https://arxiv.org/abs/quant-ph/0104088}{\tt arXiv:quant-ph/ 0104088}.

\bibitem{Paris2004}
M.~G.~A. Paris and J.~\v{R}eh\'a\v{c}ek eds., {\sl Quantum State Estimation}, Lecture Notes in Physics Vol.~649, (Springer, Berlin, 2004).

\bibitem{deFinetti1989}
B. de Finetti, ``Probabilism:\ A Critical Essay on the Theory of Probability and on the Value of Science,'' Erkenntnis {\bf 89}, 169--223 (1989).

\bibitem{Jeffrey1989}
R. Jeffrey, ``Reading {\it Probabilismo},'' Erkenntnis {\bf 89}, 225--237 (1989).

\bibitem{deFinetti2006}
B. de Finetti, {\sl L'invenzione della verita}, (Raffaello Cortina Editore, Milan, 2006).

\bibitem{James1978}
W. James, ``G. Papini and the Pragmatist Movement in Italy,'' in {\sl The Works of William James, Vol.\ 5:\ Essays in Philosophy}, (Harvard University Press, Cambridge, Massachussetts, 1978), pp.\ 144--148.

\bibitem{Fuchs12}
C. A. Fuchs, ``Interview with a Quantum Bayesian,'' in {\sl Elegance and Enigma:\ The Quantum Interviews}, edited by M.~Schlosshauer (Springer, Berlin, Frontiers Collection, 2011). \href{https://arxiv.org/abs/1207.2141}{\tt arXiv:1207.2141}.


\bibitem{Myrvold2020}
W. C. Myrvold, ``Subjectivists About Quantum Probabilities Should Be Realists About Quantum States,'' in {\sl Quantum, Probability, Logic:\ The Work and Influence of Itamar Pitowsky}, edited by M.~Hemmo and O.~Shenker, (Springer, Cham, Switzerland, 2020), pp.\ 449--465; \href{http://philsci-archive.pitt.edu/16656/}{\tt http://philsci-archive.pitt.edu/16656/}.

\bibitem{Earman2019}
J. Earman, ``Quantum Bayesianism Assessed,'' The Monist {\bf 102}, 403–423 (2019).

\bibitem{Fuchs2020}
C. A. Fuchs and B. C. Stacey, ``QBians Do Not Exist,'' \href{https://arxiv.org/abs/2012.14375}{\tt arXiv:2012.14375}.

\bibitem{Timpson08a}
C. G. Timpson, ``Quantum Bayesianism:\ A Study,'' Stud.\ Hist.\ Phil.\ Mod.\ Phys.\ {\bf 39}, 579--609 (2008); \href{https://arxiv.org/abs/0804.2047}{\tt arXiv:0804.2047}.

\bibitem{Bacciagaluppi2014}
G. Bacciagaluppi, ``A critic looks at qBism,'' in {\sl New Directions in the Philosophy of Science},'' edited by M.~C.~Galavotti, D.~Dieks, W.~Gonzalez, S.~Hartmann, T.~Uebel, and M.~Weber, (Springer, Cham, Switzerland, 2020), pp.\ 403--416; \href{http://philsci-archive.pitt.edu/9803/}{\tt http://philsci-archive.pitt.edu/9803/}.

\bibitem{Glick2021}
D. Glick, ``\href{https://link.springer.com/article/10.1007/s13194-021-00366-5}{QBism and the limits of scientific realism},'' Euro.\ J. Phil. Sci.\ {\bf 11}, 53 (2021).

\bibitem{Fuchs2019}
C. A. Fuchs and B. C. Stacey, ``Are Non-Boolean Event Structures the Precedence or Consequence of Quantum Probability?,'' \href{https://arxiv.org/abs/1912.10880}{\tt arXiv:1912.10880}.

\bibitem{Schiller1902}
F.~C.~S. Schiller, ``\href{https://en.wikisource.org/wiki/Personal_Idealism/Axioms_as_Postulates}{Axioms as Postulates},'' in {\sl Personal Idealism:\ Philosophical Essays by Eight Members of the University of Oxford}, edited by H.~Sturt, (Macmillan and Co., New York, 1902);

\bibitem{Hodgson1898}
S. H. Hodgson, {\sl The Metaphysic of Experience}, in four books, (Longmans, Green, and Co., London, 1898); reprinted by (Thoemmes Press, Bristol, England, 2001).

\bibitem{Lamberth1999}
D. C. Lamberth, {\sl William James and the Metaphysics of Pure Experience}, (Cambridge University Press, Cambridge, UK, 1999).

\bibitem{Stacey2022}
B. C. Stacey, ``Towards a World Game-Flavored as a Hawk's Wing,'' to appear in this volume; for an earlier version of the paper see, B.~C. Stacey, ``Ideas Abandoned en Route to QBism,'' \href{https://arxiv.org/abs/1911.07386}{\tt arXiv:1911.07386}.

\bibitem{FuchsStacey2018}
C. A. Fuchs and B. C. Stacey, ``QBism:\ Quantum Theory as a Hero's Handbook,'' in {\sl Proceedings of the International School of Physics ``Enrico Fermi'' Course 197 -- Foundations of Quantum Physics}, edited by E.~M. Rasel, W.~P. Schleich, and S. W\"olk (IOS Press, Amsterdam; Societ\`a Italiana di Fisica, Bologna, 2018), pp.\ 133--202; \href{https://arxiv.org/abs/1612.07308}{\tt arXiv:1612.07308}.

\bibitem{Mermin2019}
N. D. Mermin, ``Making better sense of quantum mechanics,'' Rep.\ Prog.\ Phys.\ {\bf 58}, 012002 (2019); \href{https://arxiv.org/abs/1809.01639}{\tt arXiv:1809.01639}.

\bibitem{Fuchs2002}
C. A. Fuchs, ``Quantum Mechanics as Quantum Information (and only a little more),'' \href{https://arxiv.org/abs/quant-ph/0205039}{\tt arXiv:quant-ph/0205039}.

\bibitem{Wigner1961}
E. P. Wigner, ``Remarks on the Mind-Body Question,'' in {\sl The Scientist Speculates}, edited by I. J. Good (William Heinemann, Ltd., London, 1961), pp.\ 284--302; reprinted in E. P. Wigner, {\sl Symmetries and Reflections:\ Scientific Essays of Eugene Wigner}, (Ox Bow Press, Woodbridge, CT, 1979), pp.\ 171--184.

\bibitem{Frauchiger2018}
D. Frauchiger and R. Renner, ``Quantum theory cannot consistently describe the use of itself,'' Nature Communications {\bf 9}, article 3711, (2018); \href{https://arxiv.org/abs/1604.07422}{\tt arXiv:1604.07422}.

\bibitem{Baumann2020}
V. Baumann and \v{C}. Brukner, ``Wigner's Friend as a Rational Agent,'' in {\sl Quantum, Probability, Logic:\ The Work and Influence of Itamar Pitowsky}, edited by M.~Hemmo and O.~Shenker, (Springer, Cham, Switzerland, 2020), pp.\ 91--99; \href{https://arxiv.org/abs/1901.11274}{\tt arXiv:1901.11274}.

\bibitem{Cavalcanti2021}
E. Cavalcanti, ``The View from the Wigner Bubble,'' {\sl Foundations of Physics\/} {\bf 51}, 39 (2021); \href{https://arxiv.org/abs/2008.05100}{\tt arXiv:2008.05100}.

\bibitem{Khrennikov17}
A. Khrennikov, ``\href{https://link.springer.com/article/10.1007/s1}{Towards Better Understanding QBism},'' Found.\ Sci.\ {\bf 23}, 181--195 (2018).

\bibitem{Mueller2018}
T. M\"uller and H. J. Briegel, ``\href{https://onlinelibrary.wiley.com/doi/epdf/10.1111/1746-8361.12222}{A stochastic process model for free agency under indeterminism},'' Dialectica {\bf 72}, 219--252 (2018).

\bibitem{Briegel2012}
H. J. Briegel, ``\href{https://www.nature.com/articles/srep00522}{On creative machines and the physical origins of freedom},'' Sci.\ Rep.\ {\bf 2}, 522 (2012).

\bibitem{DeBrotaStacey2018}
J. B. DeBrota and B. C. Stacey, ``FAQBism,'' \href{https://arxiv.org/abs/1810.13401}{\tt arXiv:1810.13401}.

\bibitem{DeBrota2020b}
J. B. DeBrota, C. A. Fuchs, and R. Schack, ``Respecting One's Fellow:\ QBism's Analysis of Wigner's Friend,'' Found.\ Phys.\ 50, 1859--1874 (2020); \href{https://arxiv.org/abs/2012.14375}{\tt arXiv:2008.03572}.

\bibitem{Furusawa1998}
A.~Furusawa, J.~L. S{\o}rensen, S.~L.\ Braunstein, C.~A. Fuchs, H.~J. Kimble, and E.~S. Polzik, ``\href{https://www.science.org/doi/10.1126/science.282.5389.706}{Unconditional Quantum Teleportation},'' Science {\bf 282}, 706--709 (1998).

\bibitem{Pienaar2022}
J. Pienaar, private communication, 7 December 2022.

\bibitem{Fuchs2015}
C. A. Fuchs and R. Schack, ``QBism and the Greeks:\ Why a Quantum State Does Not Represent an Element of Physical Reality,'' Phys.\ Scripta {\bf 90}, 015104 (2015); \href{https://arxiv.org/abs/1412.4211}{\tt arXiv:1412.4211}.

\bibitem{FuchsInterview}
C. A. Fuchs with C. S. Powell, ``\href{https://www.discovermagazine.com/the-sciences/quantum-physics-is-no-more-mysterious-than-crossing-the-street}{Quantum Physics is No More Mysterious Than Crossing the Street:\ A Conversation with Chris Fuchs},'' Discover Magazine, 29 November 2019.

\bibitem{Wheeler1987}
J.~A. Wheeler, ``\href{https://jawarchive.files.wordpress.com/2012/03/howcomethequantum.pdf}{How Come the Quantum?}''\ in {\it New Techniques
and Ideas in Quantum Measurement Theory}, edited by D.~M. Greenberger, Ann.\ New York Acad.\ Sci.\ {\bf 480}, 304--316 (1987).

\bibitem{Savage1972}
L. J. Savage, {\sl The Foundations of Statistics}, 2nd ed., (Dover, New York, 1972).

\bibitem{BernardoSmith1994}
J. M. Bernardo and A. F. M. Smith, {\sl Bayesian Theory}, (Wiley, Chichester, 1994).

\bibitem{deFinetti1990}
B. de Finetti, {\sl Theory of Probability}, (Wiley, New York, 1990).

\bibitem{Berkovitz2019}
J. Berkovitz, ``\href{https://doi.org/10.1007/s13194-018-0226-4}{On de Finetti's instrumentalist philosophy of probability},'' Euro.\ J. Phil.\ Sci.\ {\bf 9}, 25 (2019).

\bibitem{Schroedinger1935}
E. Schr\"odinger, ``\href{https://www.jstor.org/stable/pdf/986572.pdf}{The Present Situation in Quantum Mechanics:\ A Translation of Schr\"o\-dinger's `Cat Paradox' Paper},'' translated by J.~D. Trimmer, Proc.\ Am.\ Philos.\ Soc.\ {\bf 124}, 323--338 (1980).

\bibitem{Jaynes1990}
E. T. Jaynes, ``Probability in Quantum Theory,'' in {\sl Complexity, Entropy, and the Physics of Information}, edited by W.~H. Zurek, (Addison-Wesley, New York, 1990), pp.\ 381--403.

\bibitem{Cox1961}
R. T. Cox, {\sl The Algebra of Probable Inference}, (Johns Hopkins Press, Baltimore, MD, 1961).

\bibitem{Nielsen2010}
M. A. Nielsen and I.~L. Chuang, {\sl Quantum Computation and Quantum Information}, 10th Anniversary Edition, (Cambridge University Press, Cambridge, England, 2010).

\bibitem{Peres78}
A. Peres, ``\href{https://aapt.scitation.org/doi/pdf/10.1119/1.11393}{Unperformed Experiments Have No Results},'' Am.\ J. Phys.\ {\bf 46}, 745--747 (1978).

\bibitem{Samizdat2}
C.~A. Fuchs, {\sl My Struggles with the Block Universe:\ Selected Correspondence, January 2001 -- May 2011}, edited by Blake C. Stacey, foreword by Maximilian Schlosshauer (2014), 2,349 pages; \href{https://arxiv.org/abs/1405.2390}{\tt arXiv:1405.2390}.


\bibitem{Fuchs16}
C. A. Fuchs, ``On Participatory Realism,'' in {\sl Information and Interaction:\ Eddington, Wheeler, and the Limits of Knowledge}, edited by I.~T. Durham and D.~Rickles, (Springer, Berlin, 2016), pp.\ 113--134; \href{https://arxiv.org/abs/1601.04360}{\tt arXiv:1601.04360}.

\bibitem{Fuchs18}
C.~A. Fuchs, ``Copenhagen Interpretation Delenda Est?,''  \href{https://arxiv.org/abs/1809.05147}{\tt arXiv:quant-ph/1809.05147}.

\bibitem{Fuchs2010}
C.~A. Fuchs, \href{https://www.amazon.com/Coming-Age-Quantum-Information-Paulian/dp/0521199263}{\sl Coming of Age with Quantum Information:\ Notes on a Paulian Idea}, (Cambridge University Press, Cambridge, UK, 2010).

\bibitem{Pauli94}
W.~Pauli, {\sl Writings on Physics and Philosophy}, edited by C.~P. Enz and K. von~Meyenn, (Springer-Verlag, Berlin, 1994).

\bibitem{Pauli55}
W. Pauli, letter to Niels Bohr, dated 15 February 1955, photocopy obtained from the Niels Bohr
Institute via H. J. Folse.  More extensive excerpts from the letter can be found in Ref.~\cite{Fuchs2017}.

\bibitem{Pienaar2020}
J. Pienaar, ``Extending the Agent in QBism,'' Found.\ Phys.\ {\bf 50}, 1894--1920 (2020); \href{https://arxiv.org/abs/2004.14847}{\tt arXiv: 2004.14847}.

\bibitem{Caves07}
C.~M. Caves, C.~A. Fuchs, and R.~Schack, ``Subjective Probability and Quantum Certainty,'' Stud.\ Hist.\ Phil.\ Mod.\ Phys.\ {\bf 38}, 255--274 (2007); \href{https://arxiv.org/abs/quant-ph/0608190}{\tt arXiv:quant-ph/0608190}.

\bibitem{Mermin2014}
N. D. Mermin, ``Why QBism Is Not the Copenhagen Interpretation and What John Bell Might Have Thought of It,'' in {\sl Quantum [Un]Speakables II:\ Half a Century of Bell's Theorem}, edited by R.~Bertlmann and A.~Zeilinger, (Springer, Berlin, 2017), pp.\ 83--94; \href{https://arxiv.org/abs/1409.2454}{\tt arXiv:1409.2454}.

\bibitem{Healey2016}
R. Healey, ``\href{https://plato.stanford.edu/entries/quantum-bayesian/}{Quantum-Bayesian and Pragmatist Views of Quantum Theory},'' {\sl Stanford Encylopedia of Philosophy\/} (2016).

\bibitem{Wallace2016}
D. Wallace, ``A Case for QBism,'' presentation at the XII International Ontology Congress, San Sebastian, Spain, 5 October 2016.

\bibitem{Healey2017}
R. Healey, {\sl The Quantum Revolution in Philosophy}, (Oxford University Press, Oxford, UK, 2017).

\bibitem{Healey2022}
R. Healey, ``\href{https://link.springer.com/chapter/10.1007/978-3-030-99642-0_20}{Representation and the Quantum State},'' in {\sl Quantum Mechanics and Fundamentality}, edited by V.~Allori, (Springer, Cham, Switzerland, 2022), pp.\ 303--316.

\bibitem{Barzegar2020}
A. Barzegar, ``\href{https://doi.org/10.1007/s10701-020-00347-3}{QBism Is Not So Simply Dismissed},'' Found.\ Phys.\ (2020).

\bibitem{Brown2017}
H. R. Brown, ``\href{http://philsci-archive.pitt.edu/12978/}{The reality of the wavefunction:\ old arguments and new},'' in {\sl Philosophers Look at Quantum Mechanics}, edited by A. Cordero, (Springer, Berlin, 2019), pp.\ 63--86.

\bibitem{DeBrotaStacey2020}
J. B. DeBrota and B. C. Stacey, ``Discrete Wigner Functions from Informationally Complete Quantum Measurements,'' Phys.\ Rev.\ A {\bf 102}, 032221 (2020);\href{https://arxiv.org/abs/1912.07554}{\tt arXiv:1912.07554}.

\bibitem{DeBrota2020c}
J. B. DeBrota, C. A. Fuchs, and B. C. Stacey, ``The Varieties of Minimal Tomographically Complete Measurements,'' Int.\ J. Quant.\ Info.\ 2040005 (2020); \href{https://arxiv.org/abs/1812.08762}{\tt arXiv:1812.08762}.

\bibitem{Bohr49}
N. Bohr, ``Discussion with Einstein on Epistemological Problems in Atomic Physics,'' in {\sl Albert Einstein:\ Philosopher-Scientist}, edited by P.~A. Schilpp, (MJF Books, New York, 1970), pp.\ 201--241.

\bibitem{Teleportation}
C. H. Bennett, G. Brassard, C. Cr\'epeau, R. Jozsa, A. Peres, and W. K. Wootters, ``\href{https://journals.aps.org/prl/pdf/10.1103/PhysRevLett.70.1895}{Teleporting an Unknown Quantum State via Dual Classical and Einstein-Podolsky-Rosen Channels},'' Phys.\ Rev.\ Lett.\ {\bf 70}, 1895--1899 (1993).

\bibitem{TeleportationExperiment}
Unsigned photograph, ``\href{https://www.eurekalert.org/multimedia/680168}{Experimental Setup of Quantum Teleportation Performed in 2013},'' from ``Quantum teleportation on a chip,'' American Association for the Advancement of Science (AAAS) EurekAlert, 1 April 2015.

\bibitem{DeBrota2020}
J. B. DeBrota, C. A. Fuchs, and B. C. Stacey, ``Symmetric Informationally Complete Measurements Identify the Irreducible Difference between Classical and Quantum Systems,'' Phys.\ Rev.\ Res.\ {\bf 2}, 013074 (2020); \href{https://arxiv.org/abs/1805.08721}{\tt arXiv:1805.08721}.

\bibitem{Maudlin2016}
T. Maudlin, ``\href{http://scindeks-clanci.ceon.rs/data/pdf/0353-3891/2016/0353-38911629005M.pdf}{The Metaphysics of Quantum Theory},'' Belgrade Phil.\ Annual {\bf 29}, 5--13 (2016).

\bibitem{Schroedinger54}
E. Schr\"odinger, {\sl Nature and the Greeks and Science and Humanism}, (Cambridge University Press, Cambridge, UK, 2014).

\bibitem{Ladyman2020}
J. Ladyman, ``\href{https://plato.stanford.edu/archives/spr2020/entries/structural-realism/}{Structural Realism},'' {\sl Stanford Encylopedia of Philosophy\/} (2020).

\bibitem{Fuchs2021}
C. A. Fuchs, ``\href{https://doi.org/10.1007/s11007-020-09525-6}{Interview with Physicist Christopher Fuchs},'' with R.~P. Crease and J.~Sares, Cont.\ Phil.\ Rev.\ {\bf 54}, 541--561 (2021).

\bibitem{Callender2015}
C. Callender, private communication, Grindavik, Iceland, July 2015.

\bibitem{Fuchs2002c}
C. A. Fuchs, ``The Anti-V\"axj\"o Interpretation of Quantum
Mechanics,'' in {\sl Quantum Theory:\ Reconsideration of Foundations},
edited by A.~Khrennikov (V\"axj\"o University Press, V\"axj\"o, Sweden, 2002), pp.~99--116; \href{https://arxiv.org/abs/quant-ph/0204146}{\tt arXiv:quant-ph/0204146}.

\bibitem{Fuchs2007}
C. A. Fuchs, ``Delirium Quantum:\ Or, where I will take quantum mechanics if it will let me,'' in {\sl Foundations of Probability and Physics -- 4}, edited by G.~Adenier, C.~A. Fuchs, and A.~Yu.\ Khrennikov, AIP Conference Proceedings Vol.~889, (American Institute of Physics, Melville, NY, 2007), pp.~438--462; \href{https://arxiv.org/abs/0906.1968}{\tt arXiv:0906.1968}.

\bibitem{Fuchs2009}
C.~A. Fuchs and R.~Schack, ``From Quantum Interference to Bayesian Coherence and Back Round Again,'' in {\sl Foundations of Probability and Physics -- 5}, edited by L.~Accardi et al., AIP Conference Proceedings Vol.\ 1101, (American Institute of Physics, Melville, NY, 2009), pp.\ 260--279.

\bibitem{Zauner1999}
G. Zauner, {\sl Quantum Designs:\ Foundations of a Noncommutative Design Theory}, PhD thesis, University of Vienna, 1999; translated in Int.\ J. Quant.\ Inf.\ 9, 445--507 (2011).

\bibitem{Renes2004}
J. M. Renes, R. Blume-Kohout, A.~J. Scott, and C.~M. Caves, ``Symmetric informationally complete quantum measurements,'' J.~Math.\ Phys. {\bf 45} 2171--2180 (2004); \href{https://arxiv.org/abs/quant-ph/0310075}{\tt arXiv:quant-ph/0310075}.

\bibitem{Fuchs2017b}
C. A. Fuchs, M. C. Hoang and B. C. Stacey,  ``The SIC Question:\ History and State of Play,'' Axioms {\bf 6}, 21 (2017); \href{https://arxiv.org/abs/1703.07901}{\tt arXiv:1703.07901}.

\bibitem{Boge2022}
F. J. Boge, ``Back to Kant!\ QBism, Phenomenology, and Reality from Invariants,'' to appear in this volume; \href{http://philsci-archive.pitt.edu/21571/}{http://philsci-archive.pitt.edu/21571/}.

\bibitem{Horn1994}
R. A. Horn and C.~R. Johnson, {\sl Topics in Matrix Analysis}, (Cambridge University Press, Cambridge, UK, 1994).

\bibitem{GrasslPrivate}
M. Grassl, private communication, 26 April 2022.

\bibitem{Durt2008}
T. Durt, C. Kurtsiefer, A. Lamas-Linares and A. Ling, ``Wigner tomography of two-qubit states and quantum
  cryptography,'' Phys.\ Rev.\ A \textbf{78}, 042338 (2008); \href{https://arxiv.org/abs/0806.0272}{\tt arXiv:0806.0272}.

\bibitem{Medendorp2011}
Z. E. D. Medendorp, F.~A. Torres-Ruiz, L.~K. Shalm, G.~N.~M.\ Tabia, C.~A. Fuchs and A.~M. Steinberg, ``Experimental characterization of qutrits using symmetric informationally complete positive operator-valued measurements,'' Phys.\ Rev.\ A \textbf{83}, 051801(R) (2011); \href{https://arxiv.org/abs/1006.4905}{\tt arXiv:1006.4905}.

\bibitem{Zhao2015}
Y.-Y. Zhao, N.-K. Yu, P. Kurzy\'nski, G.-Y. Xiang, C.-F. Li and G.-C. Guo, ``Experimental realization of generalized qubit measurements based on quantum walks,'' Phys.\ Rev.\ A \textbf{91}, 042101 (2015); \href{https://arxiv.org/abs/1501.05096}{\tt arXiv:1501.05096}.

\bibitem{Appleby2017b}
M. Appleby, S.~Flammia, G.~McConnell, and J.~Yard, ``SICs and Algebraic Number Theory,'' Found.\ Phys.\ {\bf 47}, 1042--1059 (2017); \href{https://arxiv.org/abs/1701.05200}{\tt arXiv:1701.05200}.

\bibitem{Appleby2021}
M. Appleby, I.~Bengtsson, M.~Grassl, M.~Harrison, and G.~McConnell, ``SIC-POVMs from Stark units: Prime dimensions $n^2+3$,'' J. Math.\ Phys.\ {\bf 63}, 112205 (2022); \href{https://arxiv.org/abs/2112.05552}{\tt arXiv: 2112.05552}.

\bibitem{Appleby2011}
D. M. Appleby, S.~T. Flammia, and C.~A. Fuchs, ``The Lie Algebraic Significance of Symmetric Informationally Complete Measurements,'' J. Math.\ Phys.\ {\bf 52}, 022202 (2011); \href{https://arxiv.org/abs/1001.0004}{\tt arXiv:1001.0004}.

\bibitem{Appleby2015}
D. M. Appleby, C. A. Fuchs, and H. Zhu, ``Group Theoretic, Lie Algebraic and Jordan Algebraic Formulations of the SIC Existence Problem,'' Quant.\ Info.\ Comput.\ {\bf 15}, 61--94 (2015); \href{https://arxiv.org/abs/1312.0555}{\tt arXiv:1312.0555}.

\bibitem{Fuchs2011a}
C.~A. Fuchs and R.~Schack, ``A Quantum-Bayesian Route to Quantum-State Space,'' Found.\ Phys.\ {\bf 41}, 345--356 (2011); \href{https://arxiv.org/abs/0912.4252}{\tt arXiv:0912.4252}.

\bibitem{Fuchs2011b}
D. M. Appleby, {\AA}.~Ericsson, and C.~A. Fuchs, ``Properties of QBist State Spaces,'' Found.\ Phys.\ {\bf 41}, 564--579 (2011); \href{https://arxiv.org/abs/0910.2750}{\tt arXiv:0910.2750}.

\bibitem{Appleby2017}
D. M. Appleby, C. A. Fuchs, B. C. Stacey, and H. Zhu, ``Introducing the Qplex:\ A Novel Arena for Quantum Theory,''  Euro.\ Phys.\ J. D {\bf 71}, 197 (2017); \href{https://arxiv.org/abs/1612.03234}{\tt arXiv:1612.03234}.

\bibitem{DeBrota2021}
J. B. DeBrota, C.~A. Fuchs, J.~L. Pienaar, and B.~C. Stacey, ``Born's rule as a quantum extension of Bayesian coherence,''  Phys.\ Rev.\ A {\bf 104}, 022207 (2021); \href{https://arxiv.org/abs/2012.14397}{\tt arXiv:2012.14397}.

\bibitem{Stacey2021}
B. C. Stacey, ``SICs and Bell Inequalities,'' in {\sl A First Course in the Sporadic SICs}, SpringerBriefs in Mathematical Physics Vol.\ 41, (Springer, Cham, 2021), pp.\ 39--55; for an earlier version of the paper see, B.~C. Stacey, ``Is the SIC Outcome There When Nobody Looks?,'' \href{https://arxiv.org/abs/1807.07194}{\tt arXiv:1807.07194}.

\bibitem{Fuchs2022}
C. A. Fuchs, M. Olshanii, and M.~B. Weiss, ``Quantum mechanics? It's all fun and games until someone loses an $i$,'' to appear in Asian J. Phys.\ (2023); \href{https://arxiv.org/abs/2206.15343}{\tt arXiv:2206.15343}.

\bibitem{Gefter2018}
A. Gefter, ``John Archibald Wheeler \& The Problem of Multiple Observership,'' presentation at the Stellenbosch Institute for Advanced Study (STIAS), Stellenbosch, South Africa, 30 May 2018.

\bibitem{Wheeler1990}
J.~A. Wheeler, ``\href{https://jawarchive.files.wordpress.com/2012/03/informationquantumphysics.pdf}{Information, Physics, Quantum:\ the Search for
Links},'' in {\sl Proceedings of the 3rd International Symposium on
Foundations of Quantum Mechanics in the Light of New Technology},
edited by S.~Kobayashi, H.~Ezawa, Y.~Murayama, and S.~Nomura
(Physical Society of Japan, Tokyo, 1990), pp.~354--368.

\bibitem{Ralston2020}
J. P. Ralston, ``Quantum Theory without Planck’s Constant,'' Int.\ J.\ Quant.\ Found.\ {\bf 6}, 48--87 (2020); \href{https://arxiv.org/abs/1203.5557}{\tt arXiv:1203.5557}.

\bibitem{Wheeler1975}
C.~M. Patton and J.~A. Wheeler, ``Is Physics Legislated by Cosmogony?,'' in {\sl Quantum Gravity:~An Oxford Symposium}, edited
by C.~J. Isham, R.~Penrose, and D.~W. Sciama (Clarendon Press, Oxford, 1975), pp.~538--605.

\bibitem{Wheeler1988}
J.~A. Wheeler, ``\href{https://ieeexplore.ieee.org/document/5390047}{World as System Self-Synthesized by Quantum Networking},'' IBM J. Res.\ Develop.\ {\bf 32}, 4--15 (1988).

\bibitem{Caves1996}
C.~M. Caves and C.~A. Fuchs, ``Quantum Information:\ How Much
Information in a State Vector?,'' in {\sl The Dilemma of Einstein,
Podolsky and Rosen -- 60 Years Later}, edited by A.~Mann and
M.~Revzen, Ann,\ Israel Phys.\ Soc.\ {\bf 12}, 226--257
(1996); \href{https://arxiv.org/abs/quant-ph/96010125}{\tt arXiv: quant-ph/96010125}.

\bibitem{Fuchs2001}
C.~A. Fuchs, ``Quantum Foundations in the Light of Quantum
Information,'' in {\sl Decoherence and its Implications in Quantum
Computation and Information Transfer}, edited by A.~Gonis and P.~E.~A. Turchi (IOS Press, Amsterdam, 2001),
pp.\ 38--82; \href{https://arxiv.org/pdf/quant-ph/0106166.pdf}{\tt arXiv:quant-ph/ 0106166}.


\bibitem{DAriano01}
G. M. D'Ariano, P. Lo Presti, and M. G. A. Paris, ``Using Entanglement Improves the Precision of Quantum Measurements,'' Phys.\ Rev.\ Lett.\ {\bf 87}, 270404 (2001); \href{https://arxiv.org/abs/quant-ph/0109040}{\tt arXiv: quant-ph/0109040}.

\bibitem{Chiribella08}
G. Chiribella, G. M. D'Ariano, and P. Perinotti, ``Optimal Cloning of Unitary Transformation,'' Phys.\ Rev.\ Lett.\ {\bf 101}, 180504 (2008); \href{https://arxiv.org/abs/0804.0129}{\tt arXiv:0804.0129}.

\bibitem{Huelga01}
S. F. Huelga, J. A. Vaccaro, A. Chefles, and M. B. Plenio, ``Quantum remote control:\ Teleportation of unitary operations,'' Phys.\ Rev.\ A {\bf 63}, 042303 (2001); \href{https://arxiv.org/abs/quant-ph/0005061}{\tt arXiv:quant-ph/ 0005061}.

\bibitem{Richardson2006}
R. D. Richardson, {\sl William James: In the Maelstrom of American Modernism}, (Houghton Mifflin Co., Boston, 2006).

\bibitem{Fuchs2004a}
C.~A. Fuchs, R.~Schack, and P.~F. Scudo, ``A de Finetti Representation Theorem for Quantum Process Tomography,'' Phys.\ Rev.\ A {\bf 69}, 062305 (2004); \href{https://arxiv.org/abs/quant-ph/0307198}{\tt arXiv:quant-ph/0307198}.

\bibitem{Fuchs2004b}
C.~A. Fuchs and R.~Schack, ``Unknown Quantum States and Operations, a Bayesian View,'' in {\sl Quantum Estimation Theory}, edited by M.~G.~A.
Paris and J. \v{R}eh\'a\v{c}ek, (Springer-Verlag, Berlin, 2004), pp.\ 151--190; \href{https://arxiv.org/abs/quant-ph/0404156}{\tt arXiv:quant-ph/0404156}.

\bibitem{Good1983}
I. J. Good, ``46656 varieties of Bayesians,'' in {\sl Good Thinking:\ The Foundations of Probability and Its
Applications}, (Dover, Mineola, NY, 1983),  pp.\ 20–-21.

\bibitem{Jaynes2003}
E. T. Jaynes, {\sl Probability Theory:\ The Logic of Science}, (Cambridge University Press, Cambridge, UK, 2003).

\bibitem{Jeffrey1991}
R. Jeffrey, ``Introduction:\ Radical Probabilism,'' in {\sl Probability and the Art of Judgment}, (Cambridge University Press, Cambridge, UK,  1992), pp.\ 1--13.

\bibitem{Lindley2006}
D. V. Lindley, {\sl Understanding Uncertainty}, (Wiley, Hoboken, NJ, 2006).

\bibitem{Brukner01}
\v{C}. Brukner and A. Zeilinger, ``Conceptual Inadequacy of the Shannon Information in Quantum Measurements,'' Phys.\ Rev.\ A {\bf 63}, 022113 (2001); \href{https://arxiv.org/abs/quant-ph/0006087}{\tt arXiv:quant-ph/0006087}.

\bibitem{Gilton2016}
M. J. R. Gilton, ``Whence the eigenstate-eigenvalue link?,'' Stud.\ Hist.\ Phil.\ Mod.\ Phys.\ {\bf 55}, 92--100 (2016).

\bibitem{Mermin81}
N. D. Mermin, ``Quantum Mysteries for Anyone,'' J. Phil.\ {\bf 78}, 397--408 (1981).

\bibitem{PBR12}
M. F. Pusey, J. Barrett, and T. Rudolph, ``\href{https://www.nature.com/articles/nphys2309}{On the reality of the quantum state},'' Nature Phys.\ {\bf 8}, 475--478 (2012).

\bibitem{Colbeck2012}
R. Colbeck and R. Renner, ``Is a System’s Wave Function in One-to-One Correspondence with Its Elements of Reality?,'' Phys.\ Rev.\ Lett.\ {\bf 108}, 150402 (2012); \href{https://arxiv.org/abs/1111.6597}{\tt arXiv:1111.6597}.

\bibitem{Harrigan2010}
N. Harrigan and R. W. Spekkens, ``Einstein, Incompleteness, and the Epistemic View of Quantum States,'' Found.\ Phys.\ {\bf 40}, 125--157 (2010); \href{https://arxiv.org/abs/0706.2661}{\tt arXiv:0706.2661}.

\bibitem{Feynman51}
R. P. Feynman, ``The Concept of Probability in Quantum Mechanics,'' in {\sl Proceedings of the Second Berkeley Symposium on Mathematical Statistics and Probability}, edited by J.~Neyman, (University of California Press, Berkeley, 1951), pp.\ 533--541.

\bibitem{Feynman48}
R. P. Feynman, ``Space-Time Approach to Non-Relativistic Quantum Mechanics,'' Rev.\ Mod.\ Phys.\ {\bf 20}, 367--387 (1948).

\bibitem{Lindley82}
D. V. Lindley, ``Comment on A.~P. Dawid's, `The Well-Calibrated Bayesian','' J. Am.\ Stat.\ Assoc.\ {\bf 77}, 611--612 (1982).

\bibitem{Peierls91a}
R. E. Peierls, ``In defense of ‘measurement’,'' Phys.\ World {\bf 4}(1), 19--21 (1991).

\bibitem{Peierls91b}
R. Peierls, ``Observations and the `Collapse of the Wave Function,'' in {\sl More Surprises in Theoretical Physics}, (Princeton University Press, Princeton, New Jersey, 1991), 6--11.

\bibitem{Mermin01}
N. D. Mermin, ``Whose Knowledge?,'' in {\sl Quantum (Un)speakables:\ From Bell to Quantum Information}, editedby R.~Bertlmann and A.~Zeilinger (Springer, Berlin, 2001), pp.\ 271--280; \href{https://arxiv.org/abs/quant-ph/0107151}{\tt arXiv:quant-ph/0107151}.

\bibitem{BFM02}
T. A. Brun, J. Finkelstein, and N. D. Mermin, ``How much state assignments can differ,'' Phys.\ Rev.\ A {\bf 65}, 032315 (2002); \href{https://arxiv.org/abs/quant-ph/0109041}{\tt arXiv:quant-ph/0109041}.

\bibitem{CFS02}
C. M. Caves, C. A. Fuchs, and R. Schack, ``Conditions for compatibility of quantum state assignments,'' Phys.\ Rev.\ A {\bf 66}, 062111 (2002); \href{https://arxiv.org/abs/quant-ph/0206110}{\tt arXiv:quant-ph/0206110}.

\bibitem{Stacey2016}
B. C. Stacey, ``SIC-POVMs and Compatibility among Quantum States,'' Math.\ {\bf 4}, 36 (2016); \href{https://arxiv.org/abs/1404.3774}{\tt arXiv:1404.3774}.


\bibitem{Zeng2022}
V.~Baumann, \v{C}.~Brukner, E.~Cavalcanti, H.~Wiseman, and W.~Zeng, organizers, \href{https://www.wignersfriends.com/theory-workshop}{Wigner's Friends:\ Theory Workshop}, The Institute, Salesforce Tower, San Fransisco, CA, 3 Nov.\ -- 2 Dec.\ 2022.

\bibitem{Moller2008}
M. Moller, ``\href{https://williamjamesstudies.org/the-many-and-the-one-and-the-problem-of-two-minds-perceiving-the-same-thing/}{\,`The Many and the One' and the Problem of Two Minds Perceiving the Same Thing},'' William James Stud.\ {\bf 3}, (2008).

\bibitem{Greenberger2014}
D. M. Greenberger, ``\href{https://www.iqoqi-vienna.at/blogs/blog/daniel-greenberger}{Can a Computer Ever Become Conscious?},'' Vienna Quantum Cafe blog, 14 April 2014.

\bibitem{Fuchs2014}
C. A. Fuchs and R. Schack, ``Quantum Measurement and the Paulian Idea,'' in {\sl The Pauli-Jung Conjecture and Its Impact Today}, edited by H. Atmanspacher and C. A. Fuchs (Imprint Academic, Exeter, UK, 2014), pp.\ 93--107; \href{https://arxiv.org/abs/1412.4209}{\tt arXiv:1412.4209}.

\bibitem{James1882}
W. James, ``On Some Hegelisms,'' from 1882, in W.~James, {\sl The Will to Believe and Other Essays in Popular Philosophy; Human Immortality---Both Books Bound as One}, (Dover, New York, 1956), pp.\ 263--298.

\bibitem{Bell1990}
J. Bell, ``Against `measurement','' Phys.\ World {\bf 3}(8), 33--40 (1990).

\bibitem{James96a}
W.~James, {\sl Essays in Radical Empiricism}, (University of Nebraska Press, Lincoln, NB, 1996).

\bibitem{Banks2014}
E. C. Banks, {\sl The Realistic Empiricism of Mach, James, and Russell:\ Neutral Monism Reconceived}, (Cambridge University Press, Cambridge, UK, 2014).

\bibitem{Atmanspacher2022}
H. Atmanspacher and D.~Rickles, {\sl Dual-Aspect Monism and the Deep Structure of Meaning}, (Routledge, New York, 2022).

\bibitem{Plotnitsky2021}
A. Plotnitsky, {\sl Reality without Realism:\ Matter, Thought, and Technology in Quantum Physics}, (Springer, Heidelberg, 2021).

\bibitem{Wheeler83b}
J.~A. Wheeler, ``Elementary Quantum Phenomenon as Building Unit,''
in {\sl Quantum Optics, Experimental Gravity, and Measurement Theory},
edited by P.~Meystre and M.~O. Scully (Plenum Press, New York, 1983),
pp.\ 141--143.

\bibitem{Whitehead1978}
A. N. Whitehead, {\sl Process and Reality}, corrected edition, edited by D.~R. Griffin and D.~W. Sherburne, (The Free Press, New York, 1978).

\bibitem{Barad2007}
K. Barad, {\sl Meeting the Universe Halfway:\ Quantum Physics and the Entanglement of Matter and Meaning}, (Duke University Press, Durham, NC, 2007).

\bibitem{Dewey1989}
J. Dewey and A. F. Bentley, {\sl Knowing and the Known\/} (1949), in {\sl John Dewey:\ The Later Works, 1925--1953}, Vol.\ 16: 1949--1952, edited by J.~A. Boydston, (Southern Illinois University Press, Carbondale, IL, 1989).

\bibitem{Merleau-Ponty1968}
M. Merleau-Ponty, {\sl The Visible and Invisible}, edited by C.~Lefort, translated by A.~Lingis, (Northwestern University Press, Evanston, IL, 1968).

\bibitem{Barzegar2022}
A. Barzegar and D. Oriti, ``Epistemic-Pragmatist Interpretations of Quantum Mechanics:\ A Comparative Assessment,'' \href{https://arxiv.org/abs/2210.13620}{\tt arXiv:2210.13620}.

\bibitem{Brukner2022}
\v{C}. Brukner, ``\href{https://doi.org/10.1038/s42254-022-00505-8}{Wigner’s friend and relational objectivity},'' Nature Rev.\ Phys.\ {\bf 4}, 628–630 (2022).

\bibitem{Rovelli1996}
C. Rovelli, ``Relational Quantum Mechanics,'' Int.\ J. The.\ Phys. {\bf 35}, 1637--1678 (1996).

\bibitem{Brukner2018}
\v{C}. Brukner, ``A no-go theorem for observer-independent facts,'' Entropy {\bf 20}, 350 (2018); \href{https://arxiv.org/abs/1804.00749}{\tt arXiv:1804.00749}.

\bibitem{Adlam2022}
E. Adlam and C. Rovelli, ``Information is Physical:\ Cross-Perspective Links in Relational Quantum Mechanics, \href{https://arxiv.org/abs/2203.13342}{\tt arXiv:2203.13342}.

\bibitem{Pienaar2021a}
J. Pienaar, ``QBism and Relational Quantum Mechanics compared,'' Found.\ Phys. {\bf 51}, 96 (2021); \href{https://arxiv.org/abs/2108.13977}{\tt  	arXiv:2108.13977}.

\bibitem{Pienaar2021b}
J. Pienaar, ``A Quintet of Quandaries:\ Five No-Go Theorems for Relational Quantum Mechanics,'' Found.\ Phys. {\bf 51}, 97 (2021); \href{https://arxiv.org/abs/2107.00670}{\tt arXiv:2107.00670}.

\bibitem{Wheeler1982}
J.~A. Wheeler, ``Bohr, Einstein, and the Strange Lesson of the
Quantum,'' in {\sl Mind in Nature:\ Nobel Conference XVII, Gustavus
Adolphus College, St.~Peter, Minnesota}, edited by R.~Q. Elvee
(Harper \& Row, San Francisco, CA, 1982), pp.~1--23, and discussions
pp.~23--30, 88--89, 112--113, and 148--149.

\bibitem{Bertram1980}
M. Bertram, ``The Different Paradigms of Merleau-Ponty and Whitehead,'' Phil.\ Today {\bf 24}(2), 121--132 (1980).

\bibitem{Schack2022}
R. Schack, ``A QBist reads Merleau-Ponty,'' to appear in this volume; \href{https://arxiv.org/abs/2212.11094}{\tt arXiv:2212.11094}.

\bibitem{Bitbol2020}
M. Bitbol, ``A Phenomenological Ontology For Physics:\ Merleau-Ponty and QBism,'' in {\sl Phenomenological Approaches to Physics}, edited by H.~A. Wiltsche and P.~Bergofer, (Springer, Cham, 2020), pp.\ 227--242; \href{http://philsci-archive.pitt.edu/19512/}{\tt http://philsci-archive.pitt.edu/19512/}.

\bibitem{Bitbol2021}
L. de la Tremblaye and M. Bitbol, ``Towards A Phenomenological Constitution Of Quantum Mechanics:\ A QBist Approach,'' Mind and Matter {\bf 20}, 35--62 (2022).

\bibitem{Bitbol2023}
M. Bitbol and L. de la Tremblaye, ``QBism:\ An Eco-Phenomenology of Quantum Physics,'' to appear in this volume; \href{http://philsci-archive.pitt.edu/20090/}{\tt http://philsci-archive.pitt.edu/20090/}.

\bibitem{vonBaeyer2023}
H. C. von Baeyer, ``On the Consilience between QBism and Phenomenology,'' to appear in this volume; \href{https://arxiv.org/abs/2201.04734}{\tt arXiv:2201.04734}.

\end{thebibliography}
\end{document}